\documentclass[longauth]{aa} 
\usepackage{natbib}

\usepackage{graphicx}
\usepackage{subcaption} 
\usepackage{txfonts}
\usepackage{booktabs}
\usepackage{siunitx}
\usepackage{makecell}
\usepackage{orcidlink}

\begin{document} 

\title{Predictions of the Nancy Grace Roman Space Telescope Galactic Exoplanet Survey. V. Detection Rates of Multiplanetary Systems in High Magnification Microlensing Events}

\titlerunning{Roman Exoplanet Survey: Detection of Multiplanetary Systems}
\authorrunning{V. Saggese et al.}

\author{
Vito Saggese\inst{1,2,3}\,\orcidlink{0009-0000-9738-0641}
\and Étienne Bachelet\inst{4,5}\,\orcidlink{0000-0002-6578-5078}
\and Sebastiano Calchi Novati\inst{5}\,\orcidlink{0000-0002-7669-1069}
\and Valerio Bozza\inst{6,3}\,\orcidlink{0000-0003-4590-0136}
\and Giovanni Covone\inst{2,3,7}\,\orcidlink{0000-0002-2553-096X}
\and Farzaneh Zohrabi\inst{8}\,\orcidlink{0000-0003-2872-9883}
\\[2pt]\textit{(Leading Authors)}
\\[2pt]
Michael D. Albrow\inst{9}\,\orcidlink{0000-0003-3316-4012}
\and Jay Anderson\inst{10}\,\orcidlink{0000-0003-2861-3995}
\and Charles Beichman\inst{11}\,\orcidlink{0000-0002-5627-5471}
\and David P. Bennett\inst{12,13}\,\orcidlink{0000-0001-8043-8413}
\and Aparna Bhattacharya\inst{12,13}
\and Christopher Brandon\inst{14}
\and Sean Carey\inst{5}\,\orcidlink{0000-0002-0221-6871}
\and Jessie Christiansen\inst{5}\,\orcidlink{0000-0002-8035-4778}
\and Alison Crisp\inst{14}\,\orcidlink{0000-0003-4310-3440}
\and William DeRocco\inst{12,15}\,\orcidlink{0000-0003-1827-9399}
\and B. Scott Gaudi\inst{14}\,\orcidlink{0000-0003-0395-9869}
\and Jon Hulberg\inst{16,13,17}\,\orcidlink{0009-0007-3944-7298}
\and Macy J. Huston\inst{18}\,\orcidlink{0000-0003-4591-3201}
\and Stela Ishitani Silva\inst{13}\,\orcidlink{0000-0003-2267-1246}
\and Eamonn Kerins\inst{19}\,\orcidlink{0000-0002-1743-4468}
\and Somayeh Khakpash\inst{20}\,\orcidlink{0000-0002-1910-7065}
\and Katarzyna Kruszyńska\inst{21}\,\orcidlink{0000-0002-2729-5369}
\and Casey Lam\inst{22}\,\orcidlink{0000-0002-6406-1924}
\and Jessica R. Lu\inst{18}\,\orcidlink{0000-0001-9611-0009}
\and Amber Malpas\inst{14}\,\orcidlink{0000-0001-5924-8885}
\and Arjun Murlidhar\inst{14}\,\orcidlink{0009-0004-1245-092X}
\and Marz Newman\inst{8}\,\orcidlink{0009-0002-1973-5229}
\and Greg Olmschenk\inst{12,13}\,\orcidlink{0000-0001-8472-2219}
\and Matthew Penny\inst{8}\,\orcidlink{0000-0001-7506-5640}
\and Keivan G. Stassun\inst{23}\,\orcidlink{0000-0002-3481-9052}
\and Alexander P. Stephan\inst{23}\,\orcidlink{0000-0001-8220-0548}
\and Rachel A. Street\inst{21}\,\orcidlink{0000-0001-6279-0552}
\and Takahiro Sumi\inst{24}\,\orcidlink{0000-0002-4035-5012}
\and Sean K. Terry\inst{12,13}\,\orcidlink{0000-0002-5029-3257}
\and Himanshu Verma\inst{8}\,\orcidlink{0000-0002-6302-251X}
\and Weicheng Zang\inst{25}\,\orcidlink{0000-0001-6000-3463}
\\[2pt]\textit{(Roman Galactic Exoplanet Survey Project Infrastructure Team)}
}

\institute{
INAF -- IAPS, Via del Fosso del Cavaliere 100, Rome, I-00133
\and 
Dipartimento di Fisica ``Ettore Pancini'', Università di Napoli Federico II, Napoli, I-80126, Italy
\and 
INFN -- Sezione di Napoli, Via Cintia, Napoli, I-80126, Italy
\and 
Université Marie et Louis Pasteur, CNRS, Institut UTINAM UMR 6213, Besan\c{c}on, France
\and 
IPAC, Caltech, 1200 E. California Blvd., Pasadena, CA 91125, USA
\and 
Dipartimento di Fisica ``E.R. Caianiello'', Università di Salerno, Via Giovanni Paolo 132, Fisciano, I-84084, Italy
\and 
INAF -- Osservatorio Astronomico di Capodimonte, Napoli, Italy
\and 
Department of Physics and Astronomy, Louisiana State University, Baton Rouge, LA 70803, USA
\and 
University of Canterbury, Department of Physics and Astronomy, Christchurch 8020, New Zealand
\and 
Space Telescope Science Institute, 3700 San Martin Drive, Baltimore, MD 21218, USA
\and 
NASA Exoplanet Science Institute, 1200 E. California Boulevard, Pasadena, CA 91125, USA
\and 
Department of Astronomy, University of Maryland, College Park, MD 20742, USA
\and 
Code 667, NASA Goddard Space Flight Center, Greenbelt, MD 20771, USA
\and 
Department of Astronomy, The Ohio State University, Columbus, OH 43210, USA
\and 
Department of Physics \& Astronomy, The Johns Hopkins University, 3400 N. Charles Street, Baltimore, MD 21218, USA
\and 
Department of Physics, Catholic University of America, Washington, DC 20064, USA
\and 
Center for Research and Exploration in Space Science and Technology, NASA/GSFC, Greenbelt, MD 20771, USA
\and 
Department of Astronomy, University of California Berkeley, Berkeley, CA 94720, USA
\and 
Department of Physics and Astronomy, University of Manchester, Oxford Rd, Manchester M13 9PL, UK
\and 
Department of Physics, Lehigh University, 16 Memorial Drive East, Bethlehem, PA 18015, USA
\and 
Las Cumbres Observatory, 6740 Cortona Drive, Suite 102, Goleta, CA 93117, USA
\and 
Observatories of the Carnegie Institution for Science, Pasadena, CA 91101, USA
\and 
Department of Physics and Astronomy, Vanderbilt University, Nashville, TN 37235, USA
\and 
Department of Earth and Space Science, Graduate School of Science, Osaka University, Osaka, 560-0043, Japan
\and 
Department of Astronomy, Westlake University, Hangzhou 310030, Zhejiang Province, China
}

\date{Received ...}

\abstract
{The Nancy Grace Roman Space Telescope will expand the reach of gravitational microlensing surveys by increasing the number of events monitored and the precision of their light curves.}
{We investigate Roman’s ability to detect triple-lens microlensing systems, cases where a foreground star with two bound exoplanets produces detectable anomalies in a microlensing event, using its planned high-cadence observations toward the Galactic bulge.}
{We simulate a large set of high-magnification microlensing light curves based on Roman’s expected survey characteristics. A detection criterion, based on a required $\chi^2$ improvement for a two-planet model, is applied to determine whether the second planet can be reliably distinguished from a single-planet (binary-lens) model.}
{Our simulations show that the majority of two-planet microlensing events would be detectable with Roman. Events in which both planets are relatively massive (planet–star mass ratios of order $10^{-3}$), or in which the more massive planet occupies a favorable resonant configuration, produce strong central perturbations, resulting in detection efficiencies of roughly 90\%. By contrast, systems with only low-mass planets ($q \sim 10^{-4}$) or with less favorable alignments generate much weaker signals, which often fall below the detection threshold. In general, the planetary mass ratios and the resulting caustic geometry (e.g., central caustic size in resonant versus wide/close orbits) are the dominant factors governing detectability.}
{Taking into account the expected frequency of planetary systems and the fraction of high-magnification events, we estimate that Roman will detect a high-magnification triple-lens event in approximately 4.5\% of multi-planet microlensing events, corresponding to about 64 events over the course of the full survey.}


\maketitle

\section{Introduction}

In the current landscape of exoplanetary science, about six thousands of exoplanets have been discovered to date. However, only about 40\% of these are found within multiplanetary systems \citep{nasa_exoplanet_archive_ps}, i.e., systems hosting more than one planet. The majority of such systems have been discovered through the transit method \citep{1952Obs....72..199S,2003ApJ...585.1038S} and radial velocity measurements, through which the first planets around a main-sequence star was discovered \citep{1995Natur.378..355M}; with the former gaining significant success thanks to  wide-field space-based missions such as \textit{CoRot} \citep{2009A&A...506..411A}, \textit{Kepler} \citep{2010Sci...327..977B}, \textit{K2} \citep{2014PASP..126..398H} and \textit{TESS} \citep{2015JATIS...1a4003R}. These missions have revealed a rich diversity of exoplanetary architectures and enable statistical studies that, together with refined theoretical models \citep{2011ApJS..197....8L,2014ApJ...790..146F,2012A&A...547A.111M}, have challenged traditional, static models of in-situ planetary formation, prompting a paradigm shift toward more dynamic theories emphasizing planetary migration and post-formation atmospheric evolution. In addition, planet occurrence is also affected by stellar multiplicity, with tight binaries showing a suppressed planet yield \citep{2025arXiv251102643S}.
Looking ahead, the upcoming ESA mission \textit{PLATO} \citep{2024arXiv240605447R}, scheduled for launch in 2026 and designed to detect terrestrial planets in the habitable zones of Sun-like stars, will further expand the reach of transit surveys.
Among the various techniques for detecting exoplanets, gravitational microlensing plays a uniquely complementary role.  Unlike other methods, microlensing is sensitive to planets at wide orbital separations, particularly beyond the snow line, and to free-floating or non-transiting planets that would otherwise remain undetectable \citep{2023arXiv231007502M}. 
Moreover, microlensing events allow the detection of planetary systems located at kiloparsec-scale distances, far beyond the reach of most other techniques.
Statistical studies based on microlensing data \citep{2010ApJ...720.1073G,2012Natur.481..167C,2016ApJ...833..145S,2017yCat..74574089S} consistently indicate that such planets are common, with cold Neptunes and super-Earths being particularly abundant. These findings suggest that planetary systems with wide-separation planets may be the rule in our Galaxy, highlighting the importance of microlensing in constructing a complete picture of planetary demographics.
The upcoming launch of the \textit{Nancy Grace Roman Space Telescope}\footnote{\url{https://www.jpl.nasa.gov/missions/the-nancy-grace-roman-space-telescope}} \citep{2015arXiv150303757S, 2025arXiv250510574Z}  will mark a significant advancement in microlensing capabilities. Scheduled to conduct a dedicated microlensing survey toward the Galactic bulge, Roman will improve our sensitivity to distant and cold exoplanets through its wide-field infrared observations \citep{2019ApJS..241....3P}. Roman will provide data to enable an overview of planetary systems across different orbital architectures, thereby bridging the gap between close-in planets detected by transit missions and the more distant planetary populations accessible through microlensing.

To date, a total of 256 planets have been discovered through gravitational microlensing. Despite its potential, detections of multiplanetary systems via this method remain extremely rare due to the intrinsic difficulty of observation and modelling. To date, 11 planets have been discovered within binary systems \citep{2014ApJ...795...42P,2014Sci...345...46G,2016AJ....152..125B,2017AJ....154..223H,2020AJ....160...72B,2020AJ....160...64H,2021RAA....21..239Z,2021A&A...655A..24H,2021AJ....161..270H,2022MNRAS.516.1704K,2024A&A...685A..16H}
,  while fewer than a dozen microlensing-detected systems have been confirmed to host multiple planets—all of them being triple systems.
OGLE-2006-BLG-109 \citep{2008Sci...319..927G,2010ApJ...713..837B} system is the first double-planet system discovered using gravitational microlensing, with a configuration similar to the Jupiter–Saturn pair in our Solar System. The Saturn-like planet's orbital motion is even visible in the microlensing light curve, allowing partial constraints on its orbit.
OGLE-2012-BLG-0026 \citep{2013ApJ...762L..28H, 2016ApJ...824...83B} is a well-characterized system consisting of a solar-mass G-type main-sequence star in the Galactic disk, orbited by two cold gas giant planets; the masses are determined with a precision of 5\%.
OGLE-2014-BLG-1722 \citep{2018AJ....155..263S} is a likely two-planet system, consisting of a late-type host star with two cold Saturn-mass planets. It is the first multiple-planet system discovered in a low-magnification microlensing event. Even though there is no detection of parallax or finite-source effects, a Bayesian analysis constrains the planetary masses and separations.
OGLE-2018-BLG-1011 \citep{2019AJ....158..114H} is a two-planet microlensing planetary system with two Jupiter-mass planets orbiting a low-mass star. The event displays a characteristic double-peaked anomaly on its light curve. Bayesian analysis constraints the system at a distance of $\sim 7.1~\mathrm{kpc}$.
OGLE-2019-BLG0468 \citep{2022A&A...658A..93H} is a microlensing system consisting of two giant planets orbiting a G-type star with a mass of $\sim 0.9\,M_{\odot}$. The event displays a triple anomaly in its light curve, which is best explained by a triple-lens model. This system expands the known population of cold giant planets beyond the snow line.
KMT-2020-BLG-0414 \citep{2021RAA....21..239Z} is not properly a multi planetary system, it is a  triple-lens microlensing system comprising a low-mass host star orbited by an Earth-mass planet and a brown dwarf companion. The system was discovered during a high-magnification ($A_{\mathrm{max}} \sim 1450$) microlensing event, with a planet-to-host mass ratio of $q_2 \sim (0.9$–$1.2)10^{-5}$,  one of the lowest ever detected via microlensing. 
KMT-2021-BLG-1077L \citep{2022A&A...662A..70H} is a microlensing system composed of a mid-to-late M-dwarf host star orbited by two cold gas giant planets. The system lies in the Galactic bulge at a distance of $\sim 8.24~\mathrm{kpc}$, making it the most distant multi-planetary microlensing system discovered so far.
KMT-2022-BLG-1818L \citep{2025arXiv250505093L} is a microlensing system with a K-dwarf host star orbited by a super-Jupiter and a Saturn-class planet. The super-Jupiter lies on a Jupiter-like orbit, while the Saturn-class planet may be either close-in ($\sim 0.4~\mathrm{au}$) or wide ($\sim 13~\mathrm{au}$), making it in the latter case the lowest-mass planet known beyond ($10~\mathrm{au}$) in a multiplanet system.

With the advent of Roman, events of this kind are expected to become significantly more common. Roman will provide continuous, high-cadence, high-precision photometric monitoring of millions of stars, leading to a substantial increase in the detection rate of microlensing events. This will enable the detection of weaker signals from low-mass companions orbiting host stars with far greater ease than is possible with current ground-based facilities or \textit{Gaia}. Moreover, its enhanced sensitivity may even allow for the detection of exomoons.

This work is the fifth paper in the RGES yield–forecast series, which models the expected planet discoveries from the Roman microlensing survey. The first study \citep{2019ApJS..241....3P} established the simulation framework and demonstrated that, under Roman’s nominal design the survey is expected to detect roughly $1.4\times10^3$ bound exoplanets with masses above $\sim0.1\,M_{\oplus}$. Subsequent analyses extended this framework to free-floating planets \citep{2020AJ....160..123J}, predicting $\sim$250 detections down to Mars mass when adopting the planet mass function of \citet{2012Natur.481..167C}, and to exomoons orbiting wide-separation giant planets \citep{2025AJ....170..258L}, for which only of order one detection is expected over the mission lifetime. The fourth paper in the series \citep{2025arXiv251013974T} investigated the precision with which the Roman Galactic Exoplanet Survey will be able to measure host-lens masses and distances, demonstrating that the mission will meet its science requirement of achieving 20\% precision for at least 40\% of detected planet hosts.

In the present study we investigate the capability of Roman to detect microlensing events produced by triple-lens systems composed of a host star and two bound planets. We restrict our analysis to this specific configuration, as the probability of clearly detecting signals from more than three planets in a single event is extremely low due to the increased complexity of caustic structures, overlapping planetary perturbations, and the short duration of anomalies \citep{2012MNRAS.419.3631S}. Several previous works have numerically investigated the detectability of multiple-planet systems via microlensing. Early studies such as \citet{1998ApJ...502L..33G} and \citet{2001MNRAS.328..986H} demonstrated that high-magnification events offer the most favorable conditions for detecting multiple planets, due to the enhanced sensitivity to central caustic perturbations. Subsequent analyses by \citet{2005ApJ...630..535C}, \citet{2006ApJ...638.1080H} further explored the morphology and interaction of caustics in triple-lens geometries.

More recently, \citet{2023MNRAS.520.4540K} conducted a detailed numerical study of triple-lens systems modeled after a scaled Solar System analog, consisting of a star, a Jupiter-like planet, and a Saturn-like companion. They demonstrated that under certain geometrical configurations, the planetary signals remain distinguishable, especially when the caustics are well separated or partially overlapping. In parallel, \citet{2023AJ....166..140F} investigated the detectability of two-planet systems across a range of masses and separations, including events exhibiting either isolated or interacting planetary caustics; however, their analysis did not specifically focus on high-magnification events. In contrast, the present study explicitly targets high-magnification microlensing events ($A_\mathrm{max} \gtrsim 100$), which are known to provide optimal conditions for detecting low-mass and multiple planets due to the proximity of the source trajectory to the central caustic. This focus is particularly relevant in light of the expected capabilities of the \textit{Roman} mission, which will observe a substantial number of such events thanks to its high-cadence and high-precision photometric.

The structure of this paper is as follows. In Sect.~\ref{sec:Triple-Systems}, we introduce the theoretical framework of microlensing by triple-lens systems, including the relevant formalism and parametrization. In Sect.~\ref{sec:Simulations}, we describe the simulation setup, including the generation of synthetic light curves, the noise model, and the fitting procedure. In Sect.~\ref{sec:Results}, we present the results of our detectability analysis, highlighting the impact of key parameters such as mass ratios, projected separations, and source geometry. Finally, in Sect.~\ref{sec:Conclusions}, we summarize our main findings and outline future directions for extending this work.

\section{Triple-Systems Events}\label{sec:Triple-Systems}

\subsection{Principles of Microlensing} 

Gravitational microlensing occurs when a massive object (the lens) bends the light from a background source star. The multiple images of a distant star, produced by the gravitational lensing effect, cannot be resolved due to limited angular resolution. As a result, the only observable signature is a temporary increase in the star’s flux \citep{1986ApJ...304....1P,2012RAA....12..947M,2012ARA&A..50..411G,2018Geosc...8..365T, 2000A&A...357..816C,2023arXiv231007502M}. In the case of a triple-lens system, the lens consists of three point masses, such as a host star and two orbiting planets. This configuration leads to rich lensing behavior and complex light curves. The Einstein angle $\theta_E$ is a fundamental scale for the lensing configuration, defined as the angular radius of the Einstein ring. It is given by:

\begin{equation} 
\theta_E = \sqrt{\frac{4GM}{c^2} \frac{D_{LS}}{D_L D_S}}, 
\end{equation} 

where $M$ is the total mass of the lensing object, $D_L$ is the distance from the observer to the lens, $D_S$ is the distance to the source, and $D_{LS}$ is the distance from the lens to the source. Distances and positions in the lens plane are often measured in units of $\theta_E$. The lens equation relates the angular position $\vec{\theta}$ of an image on the sky to the true source position $\vec{\beta}$: 

\begin{equation} 
\vec{\beta} = \vec{\theta} - \sum_{i=0}^{2} \kappa_i \frac{\vec{\theta} - \vec{\theta}_i}{|\vec{\theta} - \vec{\theta}_i|^2}, 
\end{equation} 

where $\vec{\theta}_i$ are the angular positions of the individual lens components, and $\kappa_i$  parameters are defined as $\kappa_i = \frac{\theta_E^2 M_i}{M},$ with $M_i$ the mass of the $i$-th lens component, $M = \sum_i M_i$ the total lens mass. It is often useful to express the lens equation in complex notation \citep{1990A&A...236..311W}. Letting $z = (\theta_x + i\theta_y)/\theta_E$ and $\zeta = (\beta_x + i\beta_y)/\theta_E$ denote the complex positions in the image and source planes, respectively, and defining $z_i$ as the complex positions of the lenses, the lens equation becomes: 

\begin{equation} 
\zeta = z - \sum_{i=1}^{3} \mu_i \frac{1}{\overline{z} - \overline{z}_i}, 
\end{equation} 

where $\mu_i = M_i / \sum_j M_j$ are the normalized masses, satisfying $\sum \mu_i = 1$, and the overbar denotes complex conjugation. This compact formulation is advantageous for both analytical understanding and numerical implementation. By algebraically manipulating the complex lens equation to eliminate the complex conjugate, one obtains a single-variable complex polynomial. For an N-point-mass lens, the polynomial is of degree $N^2 + 1$. According to the image theorem \citep{2001astro.ph..3463R, 2003astro.ph..5166R, KhavinsonNeumann2006}, not all roots correspond to real images: the minimum number of images is $N + 1$, while the maximum is $5N - 5$. Therefore, a triple-lens system ($N=3$) can produce between 4 and 10 images, corresponding to the physical roots of the resulting tenth-degree polynomial. The coefficients of this polynomial depend on the normalized masses of the lenses and their complex positions. Solving this polynomial yields all possible image positions for a given source, providing a powerful method for modeling microlensing events involving three lenses.

\subsection{Magnification}

Each image produced by the lens contributes a magnification factor \( A_j \), which quantifies the local brightness enhancement of the source. For a point source, this is given by the inverse of the absolute value of the Jacobian determinant of the lens mapping \citep{1992grle.book.....S}:
\begin{equation}
A_j = \frac{1}{\left| \det J(\vec{\theta}_j) \right|}.
\end{equation}

The total magnification for a point source is obtained by summing over all images:
\begin{equation}
\mu_{\mathrm{ps}} = \sum_j A_j.
\end{equation}

For extended (i.e., finite-size) sources, the singularity at the caustic is smoothed out \citep{1994ApJ...430..505W}, and the resulting magnification peaks become finite. Assuming a uniform surface brightness and defining the normalized source radius as 
$\rho = \theta_{*} / \theta_{\mathrm{E}}$, where $\theta_{*}$ is the angular radius of the source star,
the total magnification is given by averaging the point-source magnification over the source disk:
\begin{equation}
\mu_{\mathrm{fs}}(\vec{\beta}) = \frac{1}{\pi \rho^2} \iint_{|\vec{\beta}' - \vec{\beta}| \leq \rho} \mu_{\mathrm{ps}}(\vec{\beta}') \, d^2\beta',
\end{equation}
where \( \mu_{\mathrm{ps}}(\vec{\beta}') \) is the point-source magnification at position \( \vec{\beta}' \) in the source plane.

The morphology and timescale of these smoothed magnification features depend on the source size \( \rho \) and the relative proper motion between the source and the lens, typically parameterized by the Einstein timescale \( t_{\rm E} \). The Einstein timescale is defined as $t_{\rm E} = \frac{\theta_{\rm E}}{\mu_{\rm rel}}$, where \( \mu_{\rm rel} \) is the relative proper motion between the source and the lens. It represents the time required for the source to traverse an angular distance equal to the Einstein radius, and therefore provides a direct connection between the lens mass, distances, and relative motion.  The sharp peaks that occur when the source crosses a caustic—known as caustic-crossing anomalies—encode detailed information about the lens configuration. These features are key to detecting and characterizing planetary companions.

To compute the total magnification, two main numerical techniques are commonly employed. The inverse ray-shooting method traces light rays backward from the observer through the lens plane, and estimates the magnification from the density of rays reaching the source plane. While accurate and flexible, this method is computationally intensive, especially near caustics or when modeling light curves at high resolution.

An efficient alternative is the contour integration method, which computes the magnification by integrating the inverse Jacobian along the boundaries of the lensed images in the complex plane. In this work, we adopt the publicly available \texttt{VBMicrolensing} package \citep{2025A&A...694A.219B}, which implements an optimized contour integration algorithm suitable for modeling complex events involving multiple lens components.

\subsection{Caustics and Regimes}
\label{sec:caustics}

The set of points in the source plane where $ \det J = 0 $ defines the caustics \citep{1993A&A...268..453E}. These curves play a fundamental role in gravitational microlensing, as they mark the loci where the magnification of a point source formally diverges. Physically, this divergence corresponds to a topological transition in the lens mapping, wherein two new images are created as the source enters a caustic and two images annihilate upon exiting.

In the case of binary lenses, three principal caustic topologies are commonly distinguished—\textit{close}, \textit{resonant}, and \textit{wide}—based on the projected separation $s$ (in units of $\theta_E$) between the two masses \citep{1999A&A...349..108D}. Adding a second planet to a binary lens system leads to significantly more complex caustic structures \citep{2019ApJ...880...72D}. A hierarchical triple-lens system composed of a host star and two planets typically exhibits a central caustic governed by the combined gravitational potential of all three masses, along with two planetary caustics whose locations depend on the projected separations \citep{1999A&A...348..311B, 2000A&A...359....1B}: in the wide regime, they lie near the respective planets, while in the close regime, they appear on the opposite side of the host star.

We adopt the regime classification that defines the resonant (intermediate) region based on the overlap of central and planetary caustics. Specifically, for a given mass ratio $ q $, in the asymptotic limit of $ q \to 0 $, the boundaries of the resonant regime are given by:
\begin{equation}
s_{\text{min}} = 1 - 0.75 q^{1/3}, \quad s_{\text{max}} = 1 + 1.5 q^{1/3}
\label{eq:s_min_max}
\end{equation}

These limits define three distinct regimes: \textit{Close} if $ s < s_{\text{min}} $; \textit{Resonant (Intermediate)} if $ s_{\text{min}} \leq s \leq s_{\text{max}} $; and \textit{Wide} if $ s > s_{\text{max}} $.

This definition aligns with the behavior of planetary caustics in the low-mass ratio limit ($ q \ll 1 $), where the resonant regime corresponds to configurations in which the planetary caustic merges with the central one. This merging produces a single six-cusped caustic and is known as resonant lensing, enhancing perturbations near the peak of the light curve.

To classify such configurations, the binary lens nomenclature can be naturally extended to each planet individually. Each planet is assigned to one of the three regimes based on its normalized projected separation from the host star. This results in nine possible combinations of planetary configurations in a triple-lens system.

For example, in wide–wide systems, the planetary caustics are typically well-separated and produce isolated anomalies. In contrast, configurations involving at least one planet in the resonant regime—such as close–resonant or resonant–resonant systems—can yield large, connected caustics with complex internal structure, rich in observational signatures. Moreover, the interaction between the caustics of the two planets can trigger topological transitions not present in binary lenses, such as swallow-tail structures or overlapping caustic regions, as analyzed by \citet{1998ApJ...502L..33G,2000A&A...355..423B,2001MNRAS.328..986H, 2014MNRAS.437.4006S, 2015ApJ...806...99D}.
\subsection{Parametrization}
\label{sec:parametrization}

\begin{figure}
    \centering
    \includegraphics[width=0.45\textwidth]{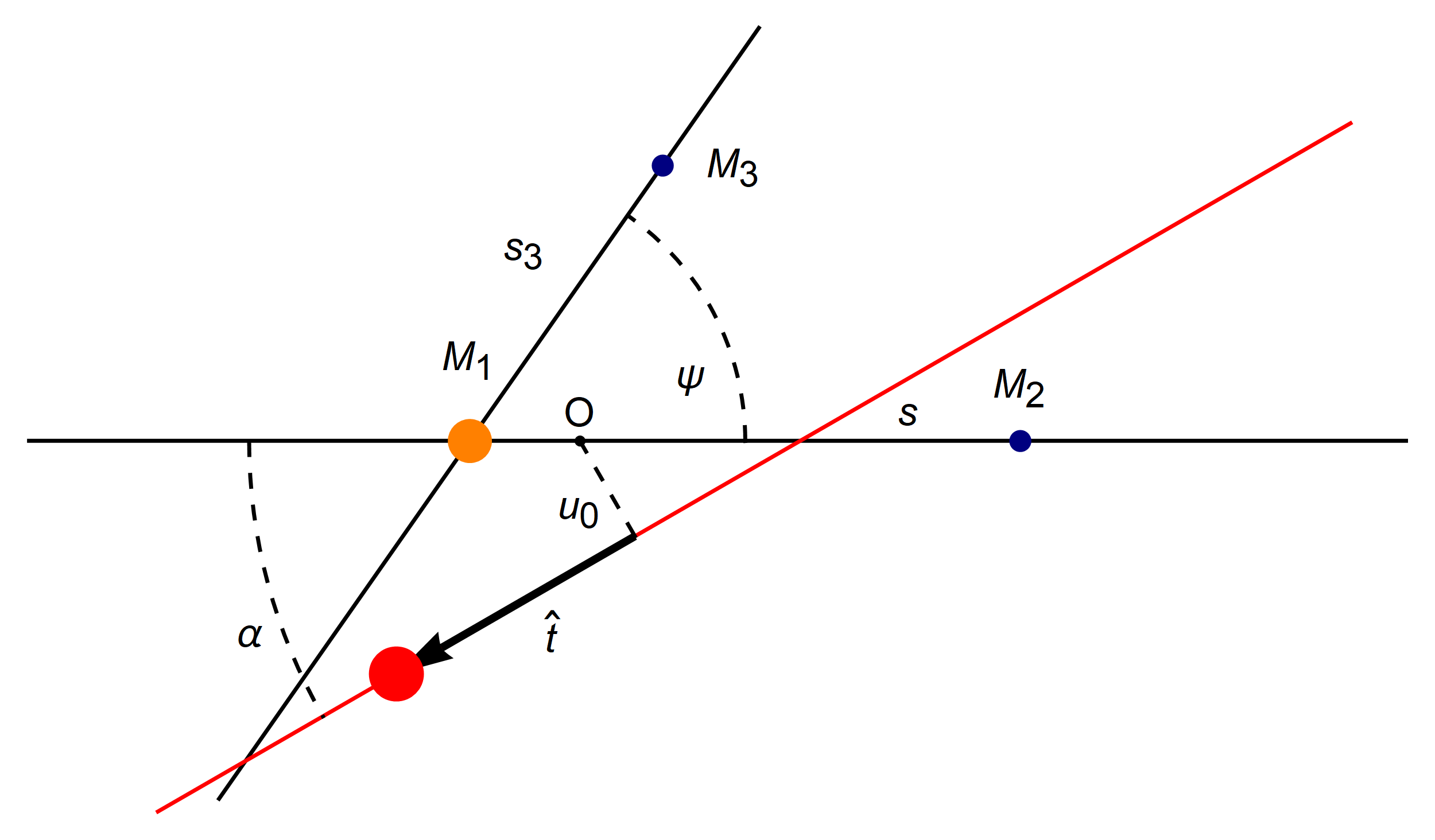}
    \caption{Schematic convention used for the geometry of triple-lens configurations. The parameters shown in the figure are defined in Section~\ref{sec:parametrization}.}
    \label{fig:sourcetrajconvention}
\end{figure}

Figure~\ref{fig:sourcetrajconvention} illustrates the geometric convention adopted for the triple-lens configuration analyzed in this work. The system consists of a primary lens (the host star) and two planetary companions. Separations are expressed in units of the Einstein angle, $ \theta_E $, and projected onto the lens plane.

The coordinate origin $ O $ is placed at the center of mass of the host star and the first planet.

The source trajectory is described by three standard microlensing parameters:
\begin{itemize}
    \item $ t_0 $: time of closest approach between the source and the lens system's center of mass;
    \item $ u_0 $: minimum impact parameter (in units of $\theta_E$);
    \item $ t_E $: Einstein ring crossing time.
\end{itemize}

To fully describe the lens system, we introduce the following parameters:
\begin{itemize}
    \item \( q \) and \( q_3 \): mass ratios of the two planets with respect to the host star;
    \item \( s \) and \( s_3 \): projected separations of each planet from the host star (in units of \( \theta_E \));
    \item \( \psi \): angle between the star–planet 1 axis and the star–planet 2 axis, measured counterclockwise.
\end{itemize}

We define \( \hat{t} \) as the unit vector parallel to, and oriented along, the source trajectory, indicating its direction of motion in the lens plane. The orientation of the source trajectory relative to the lens configuration is described by the angle \( \alpha \), defined as the angle between the source trajectory and the star–planet 1 axis.

Compared to the binary-lens case—which requires a single mass ratio \( q \), a separation \( s \), and a trajectory angle \( \alpha \)—the triple-lens parametrization introduces three additional degrees of freedom: a second mass ratio \( q_3 \), a second projected separation \( s_3 \), and the angle \( \psi \).

\section{Simulations}\label{sec:Simulations}
\subsection{Sample}

To simulate realistic triple-lens microlensing events, we built upon the Galactic microlensing population framework introduced by \citet{2019ApJS..241....3P}, employing an updated, newly generated set of simulations made available to the authors. This dataset consists of binary-lens microlensing events produced by a host star and a single planetary companion, characterized by a planet-to-host mass ratio \( q \) and a projected separation \( s \).
Triple-lens systems were constructed by pairing two binary events with host stars of similar mass, such that the fractional difference in host star masses satisfies $\left| \Delta M_\star / M_\star \right| < 0.05$. The first event provided the base configuration (\( q, s \)), while the second contributed a second planet (\( q_3, s_3 \)). The relative angular orientation \( \psi \) between the two planets was randomly sampled using 12 values distributed in the range \([0^\circ, 360^\circ]\), ensuring a broad range of projected configurations.
In all configurations, the two planets are ordered such that $q_2 > q_3$.
Although this construction method does not ensure dynamical stability, it offers an effective framework for investigating triple-lens microlensing signatures in high-magnification events.

To focus on central caustic features, we set high-magnification source trajectories with small impact parameters: \( u_0 = 0.01, 0.001, \text{and } 0.0001 \). For each triple-lens configuration, we varied the trajectory angle \( \alpha \), measured between the source path and the host--planet~1 axis. To ensure consistency with the binary-lens analysis, we adopted the same two reference values of \( \alpha \) used in the \cite{2019ApJS..241....3P} simulations.

\begin{figure*}
    \centering
    \includegraphics[width=0.95\textwidth]{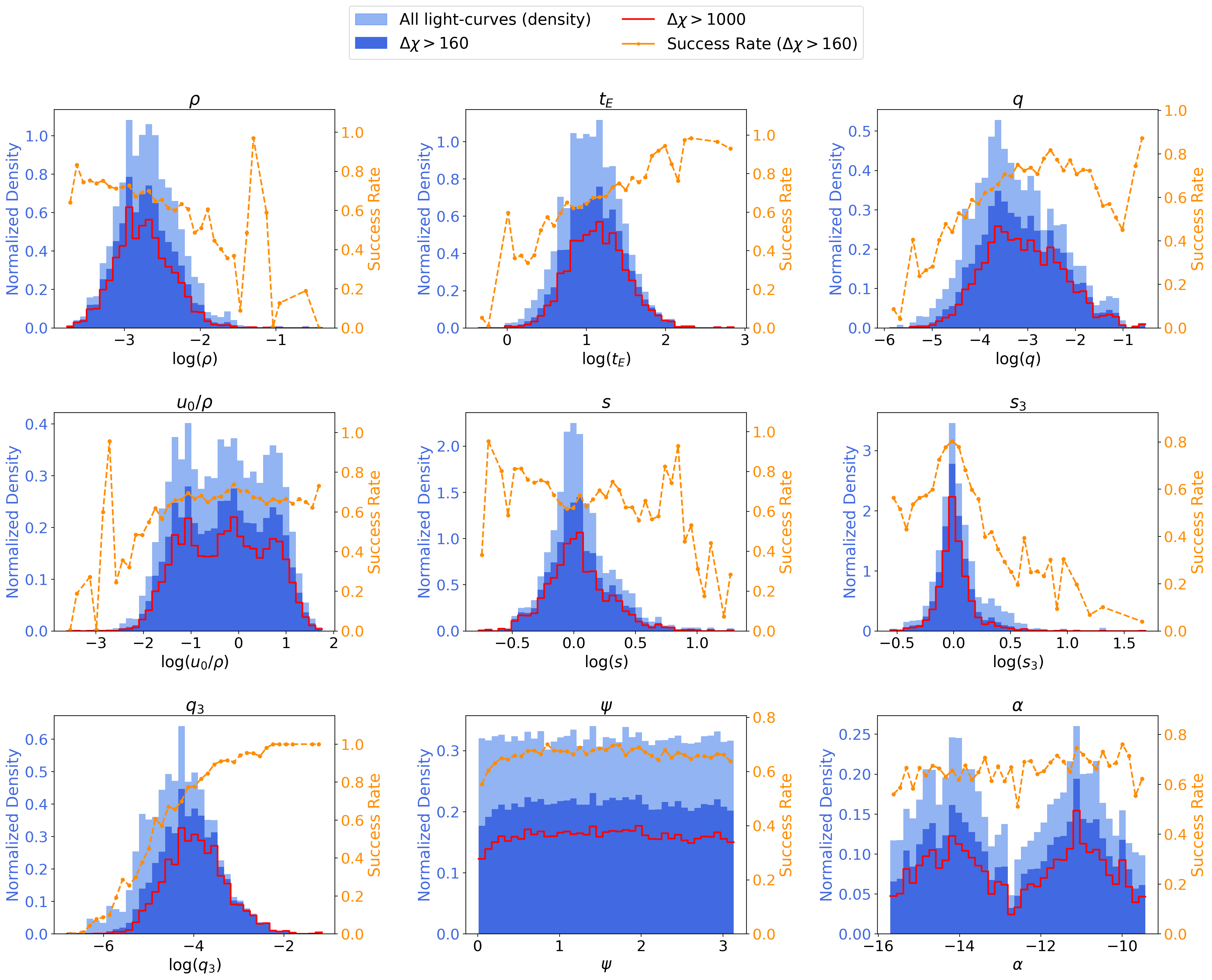}
    \caption{
    Detection efficiency of triple-lens microlensing events as a function of various physical and geometric parameters. Each panel shows the normalized distribution of all simulated events (light blue), events with significant signals ($\Delta\chi^2 > 160$, dark blue), events with very strong signals ($\Delta\chi^2 > 1000$, red line), and the corresponding detection success rate (orange curve). The efficiency increases with the outer planet’s mass ratio $q_3$, its projected separation $s_3$, and favorable source-caustic configurations.
    }
    \label{fig:success_rate_histograms}
\end{figure*}

Each model assumed a source radius \( \rho \) and Einstein timescale \( t_E \) taken from the original binary configuration. The resulting dataset consists of 1298283 synthetic light curves covering a wide variety of triple-lens geometries and observing conditions. The parameter distributions are illustrated by the light blue histogram in Figure \ref{fig:success_rate_histograms}.

\subsection{Method}

Each configuration is defined by the parameters $(q, s)$ and $(q_3, s_3)$ of the two planets, the orientation angle $\psi$, the impact parameter $u_0$, and the Einstein timescale $t_E$ and normalized source radius $\rho$.

Magnification curves were computed using \texttt{VBMicrolensing}, employing the \texttt{Multipoly} algorithm optimized for triple-lens systems. Each event was simulated over a 72-day observing window with a cadence of 15 minutes, consistent with the planned Roman survey \citep{2019ApJS..241....3P, 2025arXiv250510574Z}.

To generate the observed fluxes, we combined the theoretical magnification with the photon flux expected from the source, blend, and background stars, assuming the fluxes from the Penny simulation sample and using the W149 bandpass.  While Roman's instrumental properties are well characterized, we adopted a simplified noise model to isolate the detectability of triple-lens features. Specifically, we applied Poisson noise to the simulated fluxes, drawing photon counts from a Poisson distribution at each time step.This model neglects additional detector effects such as read noise or background, but is sufficient to capture the primary photometric limitations relevant for assessing the signal strength of triple-lens signatures in high-magnification events.

To evaluate whether the presence of a second planet (i.e., the triple-lens nature of the system) could be identified, each light curve was fit with a simpler binary-lens model. The fit was performed using a non-linear least-squares minimization based on the Levenberg–Marquardt algorithm, with bounded initial conditions derived from the true configuration. We extracted the best-fit parameters and computed the binary-lens $\chi^2_{\mathrm{binary}}$, which was then compared to the corresponding $\chi^2_{\mathrm{triple}}$ value obtained by evaluating the same noise-affected light curve under the original triple-lens model (without re-fitting). In this way, the comparison isolates the improvement in fit quality attributable to the presence of the second planet. A detection was considered significant when the difference exceeded a conservative threshold of $\Delta \chi^2 = 160$, consistent with previous microlensing detection criteria \citep[e.g.,][]{2010ApJ...720.1073G, 2018AJ....155..263S, 2019ApJS..241....3P}.

The simulation pipeline described above enables a systematic exploration of triple-lens microlensing events across a broad range of astrophysical and geometrical configurations. By varying key parameters the framework captures the diversity of light curve morphologies that can arise in multi-planet systems. This approach allows us to probe the sensitivity of microlensing detections to the underlying lens architecture and to quantify the specific regimes in which the presence of a second planet produces statistically significant deviations from simpler binary-lens models. The resulting simulation forms the methodological basis for assessing the detectability of complex planetary configurations with Roman, and provides a direct numerical complement to previous analytical studies and detectability forecasts for multiple-planet microlensing events \citep[e.g.,][]{2010ApJ...720.1073G, 2018AJ....155..263S, 2023AJ....166..140F}.

\begin{table*}[ht]
\caption[]{Detection statistics by orbital regime}
\label{tab:regime_stats}
\centering
\begin{tabular}{lcccccc}
\hline \hline
Orbital Regime &
Simulated &
Success &
Fraction of &
Detection &
Expected \\
&
Events &
Rate (\%) &
Simulations &
Probability (\%) &
Events \\
\hline
$s$ Resonant / $s_3$ Resonant     & 14\,518   & 92.82 & 0.01 & 9.32$\times 10^{-4}$ & 0.76 \\
$s$ Close / $s_3$ Resonant        & 72\,472   & 91.37 & 0.06 & 4.58$\times 10^{-3}$ & 3.73 \\
$s$ Wide / $s_3$ Resonant         & 77\,891   & 89.56 & 0.06 & 4.83$\times 10^{-3}$ & 3.93 \\
$s$ Resonant / $s_3$ Close        & 49\,281   & 80.81 & 0.04 & 2.75$\times 10^{-3}$ & 2.24 \\
$s$ Close / $s_3$ Close           & 219\,303  & 70.23 & 0.17 & 1.06$\times 10^{-2}$ & 8.68 \\
$s$ Wide / $s_3$ Close            & 263\,649  & 69.02 & 0.20 & 1.26$\times 10^{-2}$ & 10.25 \\
$s$ Resonant / $s_3$ Wide         & 46\,240   & 63.91 & 0.04 & 2.05$\times 10^{-3}$ & 1.67 \\
$s$ Close / $s_3$ Wide            & 258\,603  & 55.44 & 0.20 & 9.91$\times 10^{-3}$ & 8.08 \\
$s$ Wide / $s_3$ Wide             & 294\,748  & 54.65 & 0.23 & 1.11$\times 10^{-2}$ & 9.08 \\
\hline
\end{tabular}
\end{table*}

\begin{figure*}
    \centering
    \includegraphics[width=0.9\textwidth]{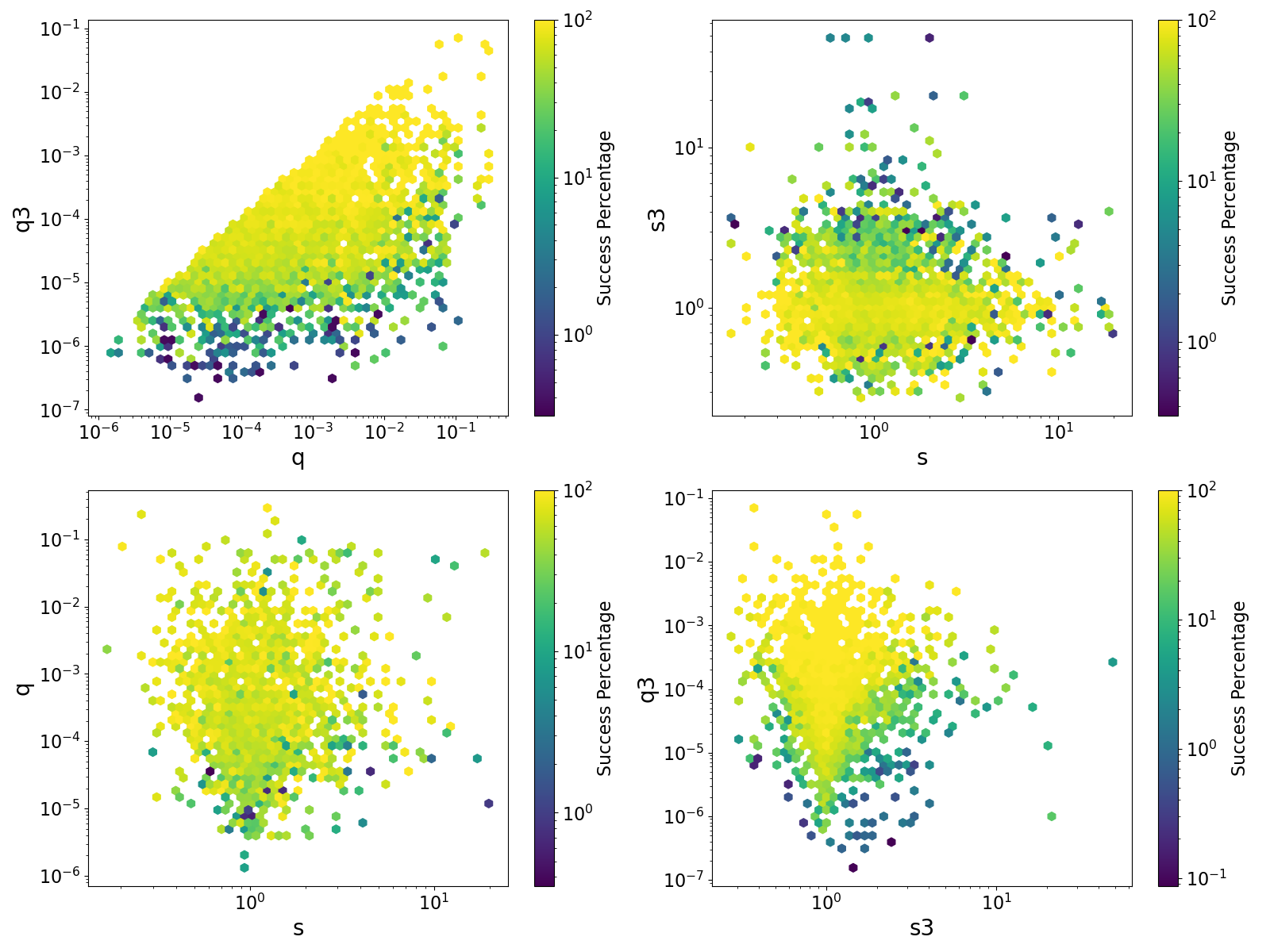}
   \caption{Detection efficiency as a function of planetary mass ratios. The color-coded map shows the fraction of detectable triple-lens events in each $(q_1, q_2)$ bin, where $q_1$ and $q_2$ are the mass ratios of the two planetary companions. Detection probability is highest when both planets are relatively massive ($q \gtrsim 10^{-3}$), while systems with low-mass companions ($q \lesssim 10^{-4}$) are generally undetectable. This trend reflects the dependence of caustic size and light curve anomalies on planetary mass.}

    \label{fig:success_rate}
\end{figure*}

\section{Results}\label{sec:Results}

In this section, we present the outcomes of our simulations aimed at evaluating Roman's ability to uncover triple-lens microlensing systems. By analyzing the ensemble of synthetic light curves under a variety of physical and observational conditions, we quantify how often the signature of a second planetary companion can be confidently recovered.
These results offer a statistical view of the parameter space where triple-lens signatures are most likely to stand out.

\subsection{Global Detection Efficiency and Orbital Regime Dependence}
\label{sec:results_overview}

Out of a total of $\approx 1.3 \times 10^6$ simulated light curves, we find that 66.3\% satisfy our detection criterion, defined as a model fit improvement of $\Delta\chi^2 > 160$ when moving from a binary-lens to a triple-lens model. This highlights the strong sensitivity of the Roman survey to deviations introduced by multiple planets, especially when their caustic structures interact.

To explore how detectability depends on orbital configuration, we classify each event according to the projected separations $s$ and $s_3$ of the two planets relative to the Einstein radius, assigning each planet to one of three regimes: \textit{close}, \textit{resonant}, or \textit{wide}. This classification is based on the mass-ratio-dependent boundaries described in Section~\ref{sec:caustics}.

Table~\ref{tab:regime_stats} presents the updated statistics using this refined classification. The results show that, in high-magnification geometries, detection efficiency is highly sensitive to whether either planet resides within the resonant zone. The highest success rates are obtained in the \textit{$s$ Resonant / $s_3$ Resonant} (92.8\%), \textit{$s$ Close / $s_3$ Resonant} (91.4\%), and \textit{$s$ Wide / $s_3$ Resonant} (89.6\%) configurations — all involving at least one planet near the Einstein radius. These cases benefit from the enhanced probability that the planetary caustic overlaps with the central magnification region, particularly during the light curve peak.

In contrast, configurations where both planets lie in the wide regime (e.g., \textit{$s$ Wide / $s_3$ Wide}) yield lower detection rates (54.7\%), reflecting the diminished central caustic influence of wide-separation planets. Intermediate success rates are observed in configurations with non-resonant separations, such as \textit{$s$ Resonant / $s_3$ Close} (80.8\%) and \textit{$s$ Close / $s_3$ Close} (70.2\%).

These findings confirm the dominant role of resonant caustics in enabling planetary detections in multi-lens events. Since the caustic structure is strongly dependent on both $s$ and $q$, the regime classification offers a more accurate framework to interpret detection statistics.

To place our detection efficiencies in a realistic context, we estimate the probability of detecting a high-magnification triple-lens microlensing event.
According to the predictions of \cite{2019ApJS..241....3P}, the \textit{Roman Space Telescope} is expected to observe approximately 54,000 microlensing events during its nominal survey, which consists of six 72-day observing seasons. Of these, about 1,400 are predicted to involve binary-lens systems with bound planetary companions, corresponding to roughly 2.6\% of the total event sample.
Based on our analysis of the Penny et al. sample, we find that 6.9\% of these binary-lens planetary events are expected to exhibit high-magnification geometries, defined as those with $u_0 < 0.01$. For such configurations, our simulations yield an average detection efficiency of 66.3\% for successfully identifying triple-lens features.
Applying these factors, assuming that all simulated high-magnification binary events host an additional planet, results in an estimated detection rate of triple-lens events of approximately 4.5\%, which, when multiplied by 1,400 events, corresponds to about 64 events over the course of the Roman survey.
The final column of Table~\ref{tab:regime_stats} refines this estimate for each orbital regime. Specifically, we compute the regime-dependent detection probability as the product of: (1) the probability that the event involves at least one bound planet, (2) the probability of a high-magnification configuration, (3) the regime-specific detection efficiency derived from our simulations, and (4) a normalization factor based on the relative frequency of each regime within the simulated triple-lens population.

\subsection{Light Curve Morphologies Across Orbital Regimes}

\label{sec:orbi_regimes}
\begin{figure*}[p]
    \centering
    \begin{subfigure}{0.49\textwidth}
        \includegraphics[width=\linewidth]{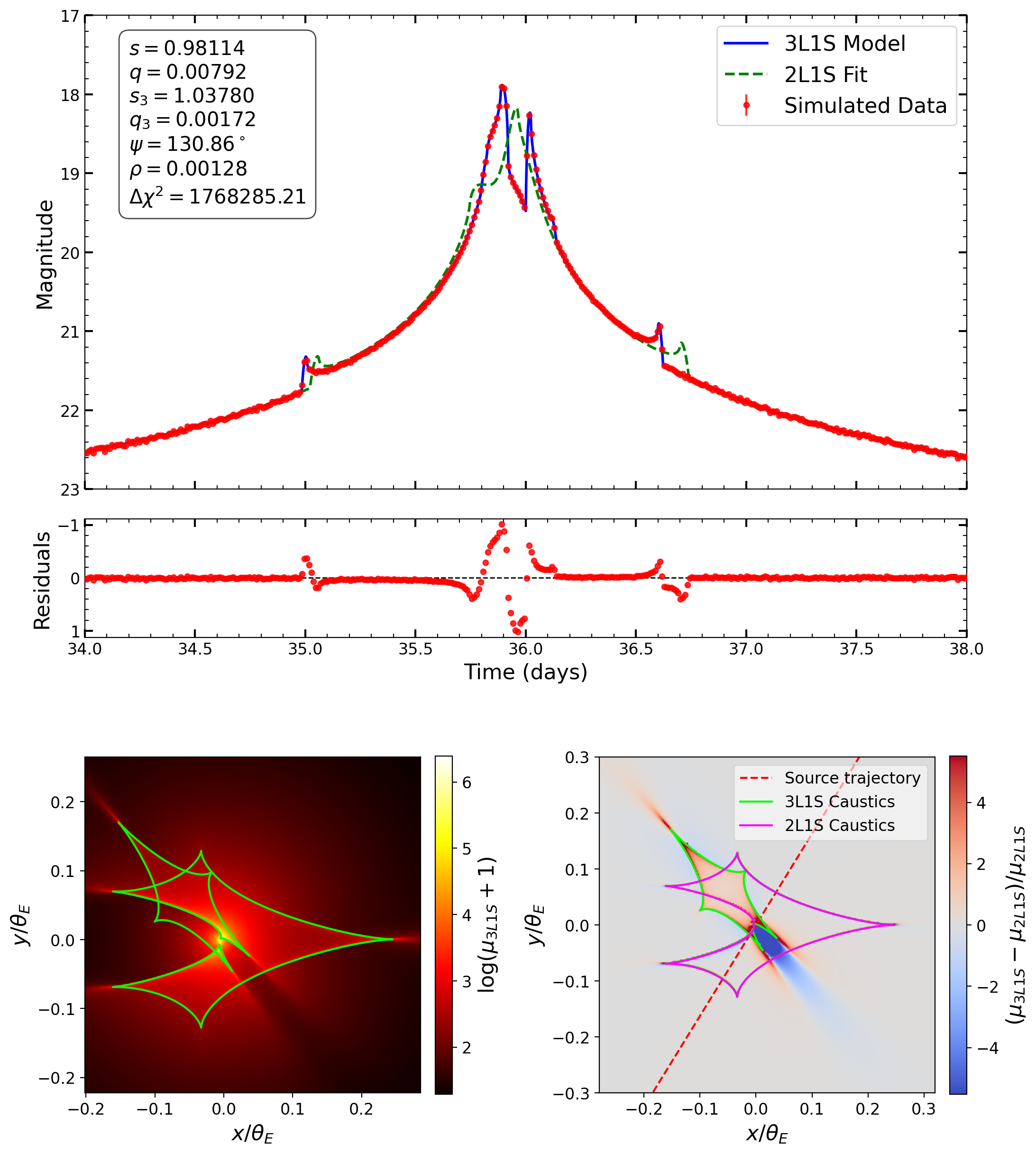}
        \caption{$s$ Resonant / $s_3$ Resonant}
    \end{subfigure}
    \hfill
    \begin{subfigure}{0.49\textwidth}
        \includegraphics[width=\linewidth]{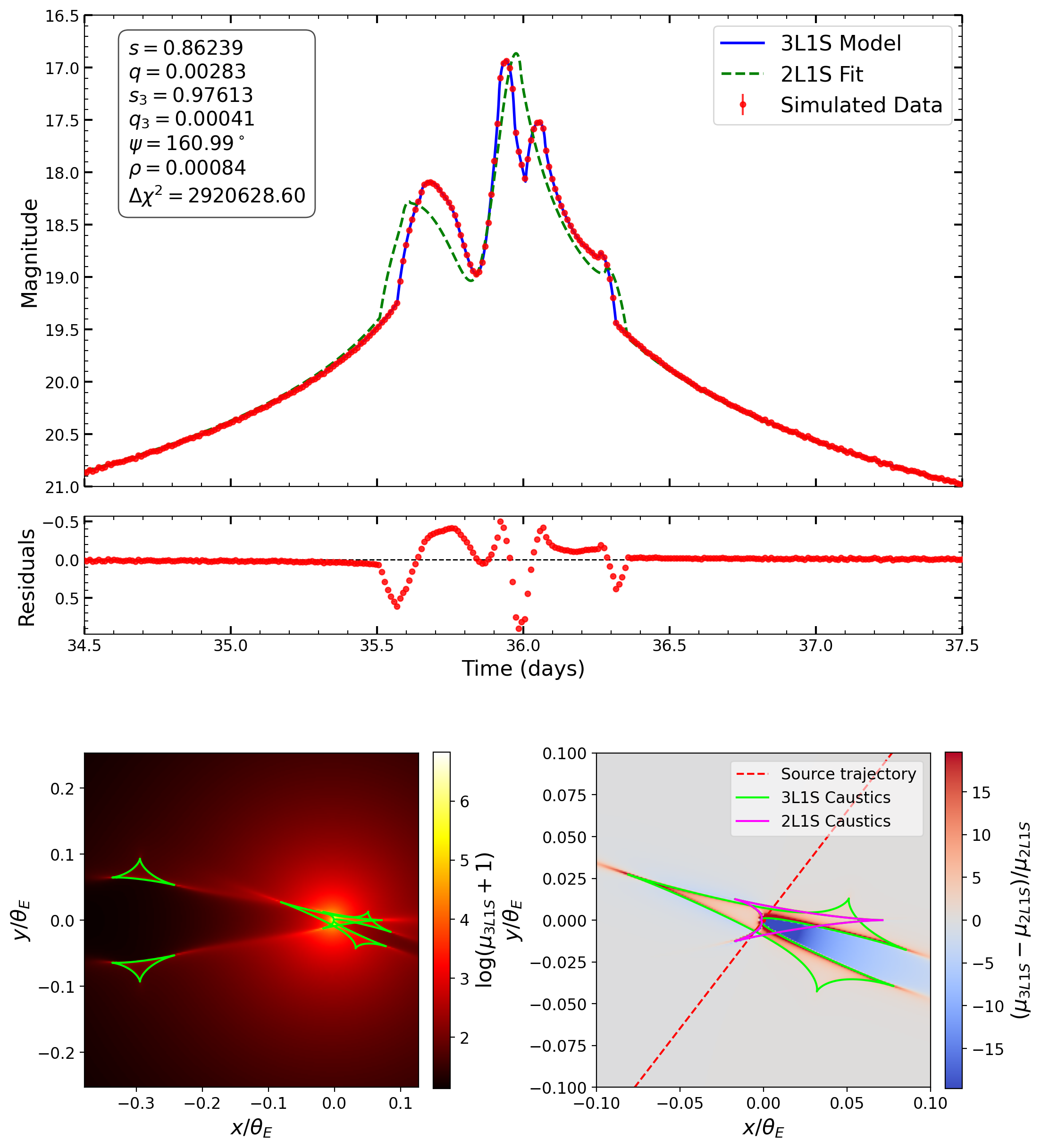}
        \caption{$s$ Close / $s_3$ Resonant}
    \end{subfigure}

    \vspace{1em}

    \begin{subfigure}{0.49\textwidth}
        \includegraphics[width=\linewidth]{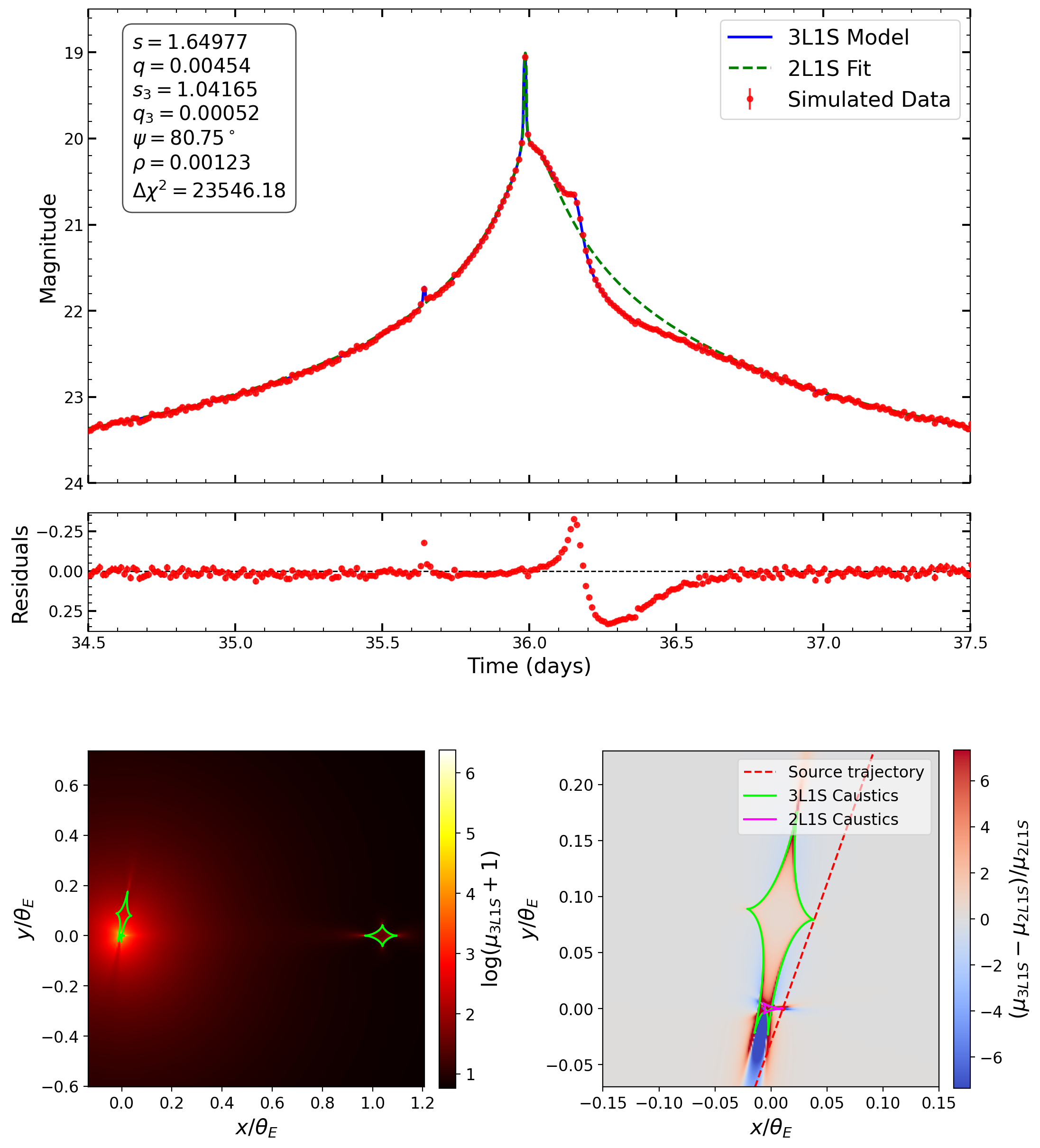}
        \caption{$s$ Wide / $s_3$ Resonant}
    \end{subfigure}

    \caption{Simulated light curves for triple-lens systems where at least one planet lies in the resonant regime, as defined by the mass-ratio-dependent boundaries (Equations~\ref{eq:s_min_max}) These configurations generally show strong central caustic perturbations and yield the highest detection efficiencies in our sample. For each case, the two smaller panels show the triple-lens magnification map and the map of the relative difference between the triple- and binary-lens models near the central caustic. }
    \label{fig:resonant_regimes}
\end{figure*}

\begin{figure*}[p]
    \centering
    \begin{subfigure}{0.49\textwidth}
        \includegraphics[width=\linewidth]{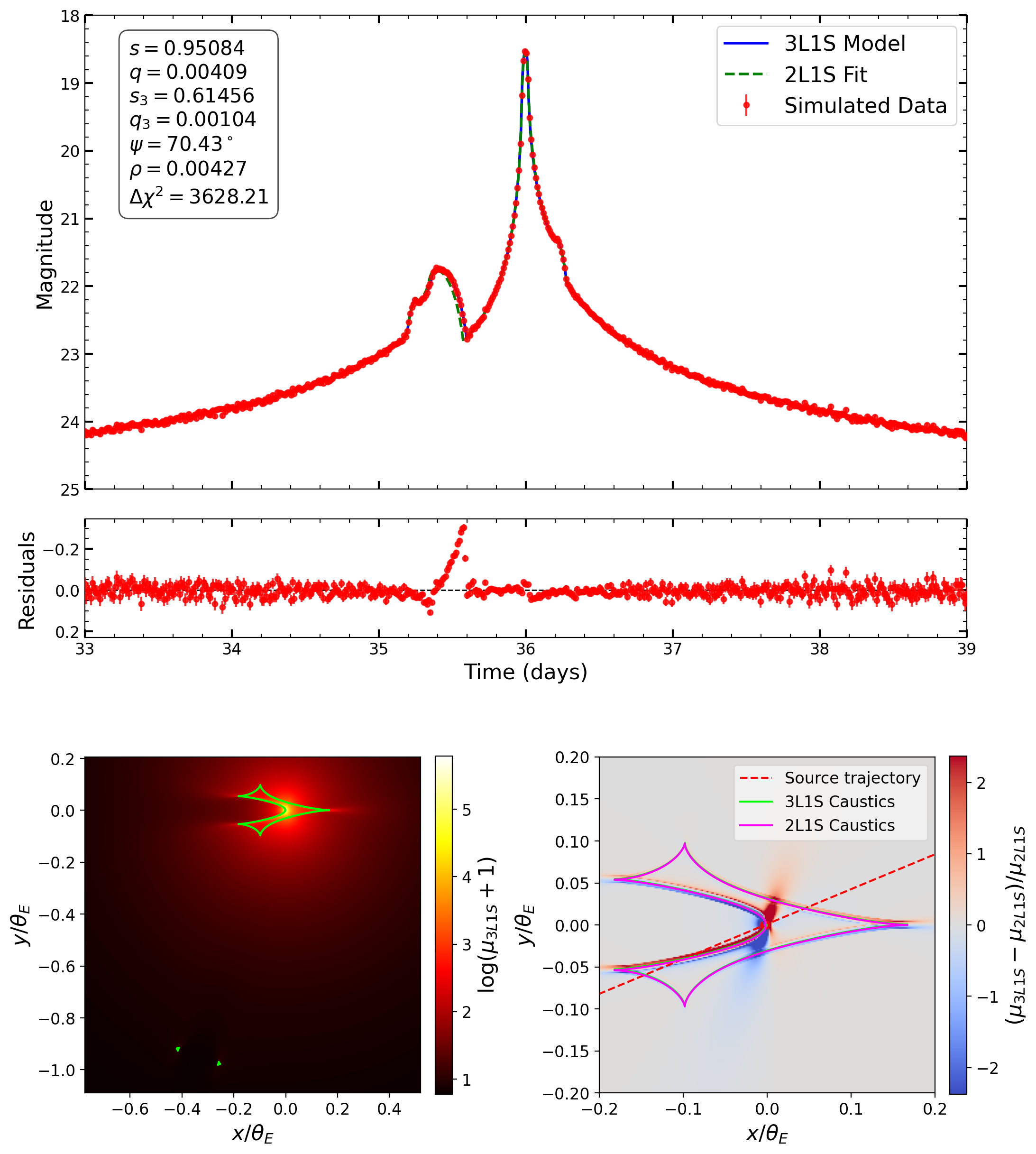}
        \caption{$s$ Resonant / $s_3$ Close}
    \end{subfigure}
    \hfill
    \begin{subfigure}{0.49\textwidth}
        \includegraphics[width=\linewidth]{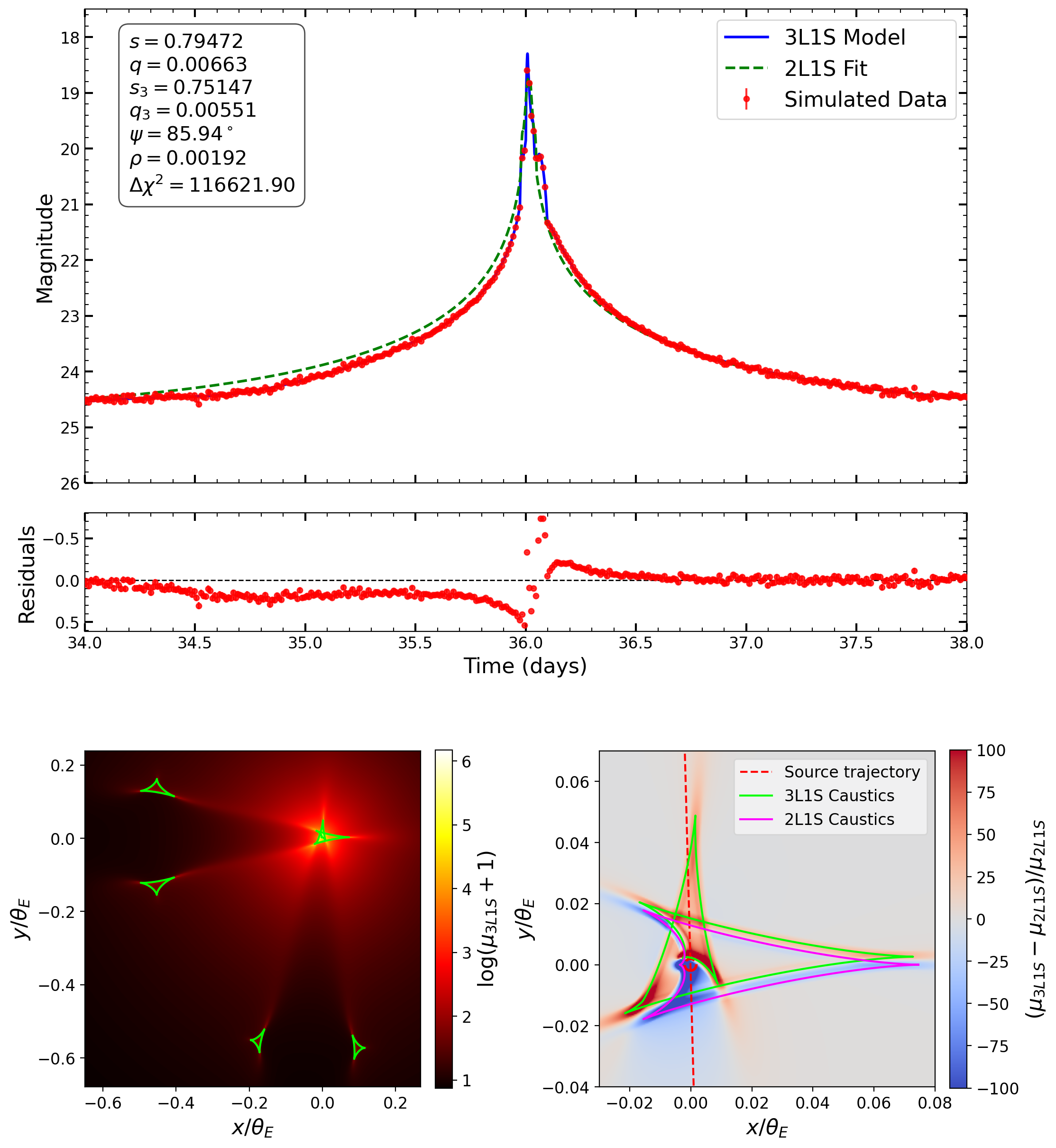}
        \caption{$s$ Close / $s_3$ Close}
    \end{subfigure}

    \vspace{1em}

    \begin{subfigure}{0.49\textwidth}
        \includegraphics[width=\linewidth]{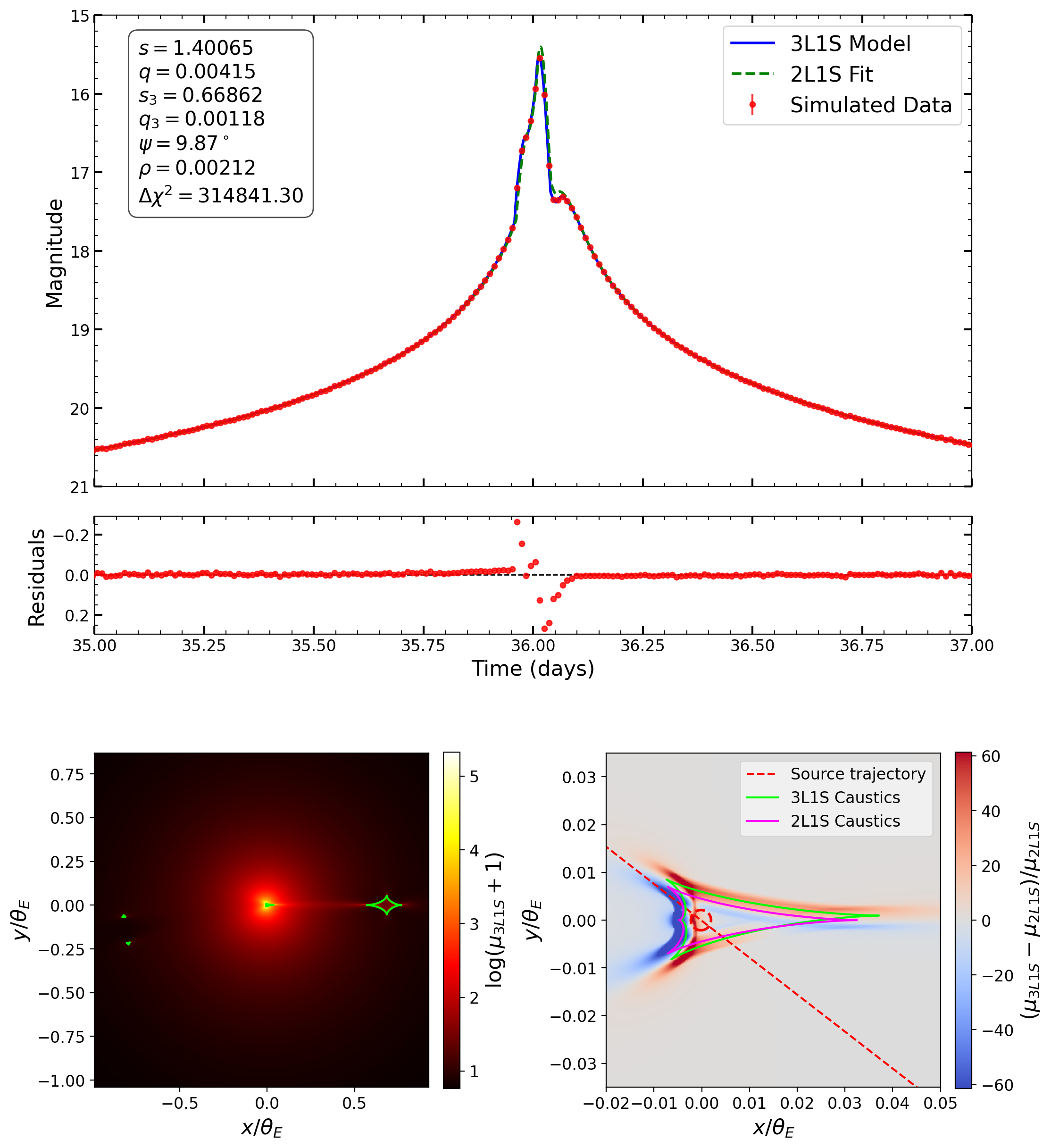}
        \caption{$s$ Wide / $s_3$ Close}
    \end{subfigure}

    \caption{Simulated light curves for configurations involving close-separation planets, or a resonant–close mix. These regimes produce intermediate detection efficiencies, with caustics that are typically more compact or asymmetric than in resonant-resonant cases.  For each case, the two smaller panels show the triple-lens magnification map and the map of the relative difference between the triple- and binary-lens models near the central caustic.}
    \label{fig:close_regimes}
\end{figure*}

\begin{figure*}[p]
    \centering
    \begin{subfigure}{0.49\textwidth}
        \includegraphics[width=\linewidth]{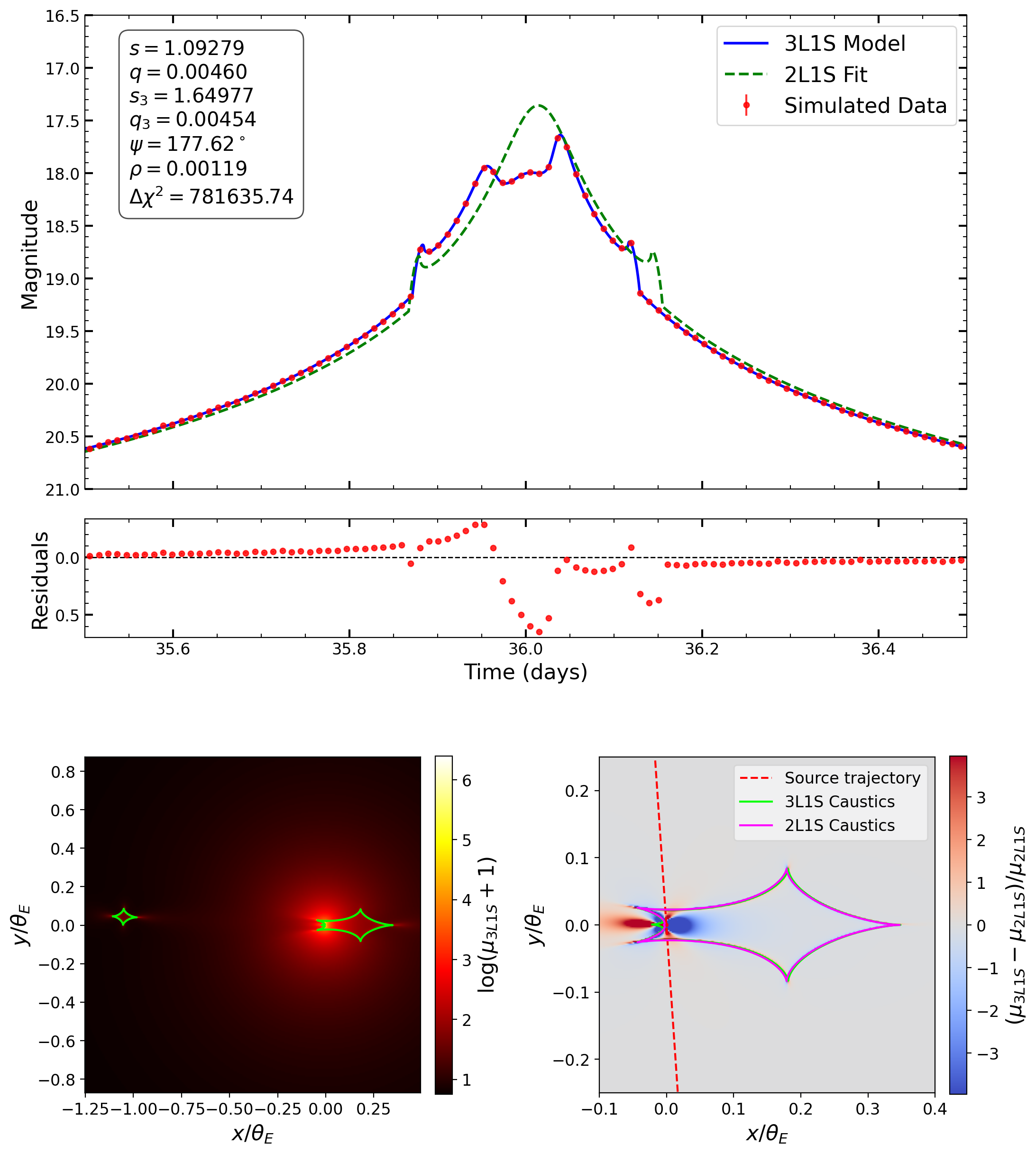}
        \caption{$s$ Resonant / $s_3$ Wide}
    \end{subfigure}
    \hfill
    \begin{subfigure}{0.49\textwidth}
        \includegraphics[width=\linewidth]{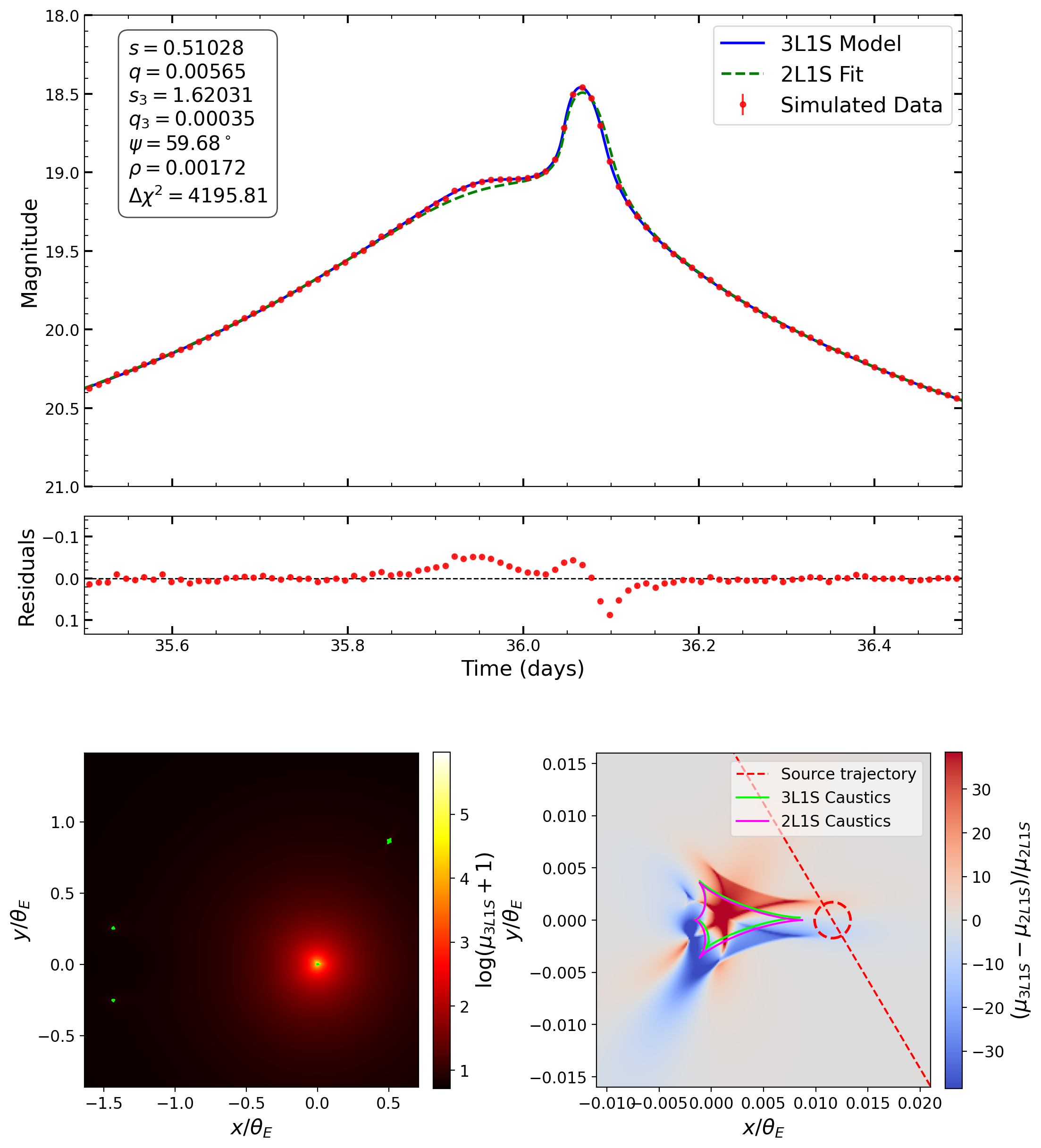}
        \caption{$s$ Close / $s_3$ Wide}
    \end{subfigure}

    \vspace{1em}

    \begin{subfigure}{0.49\textwidth}
        \includegraphics[width=\linewidth]{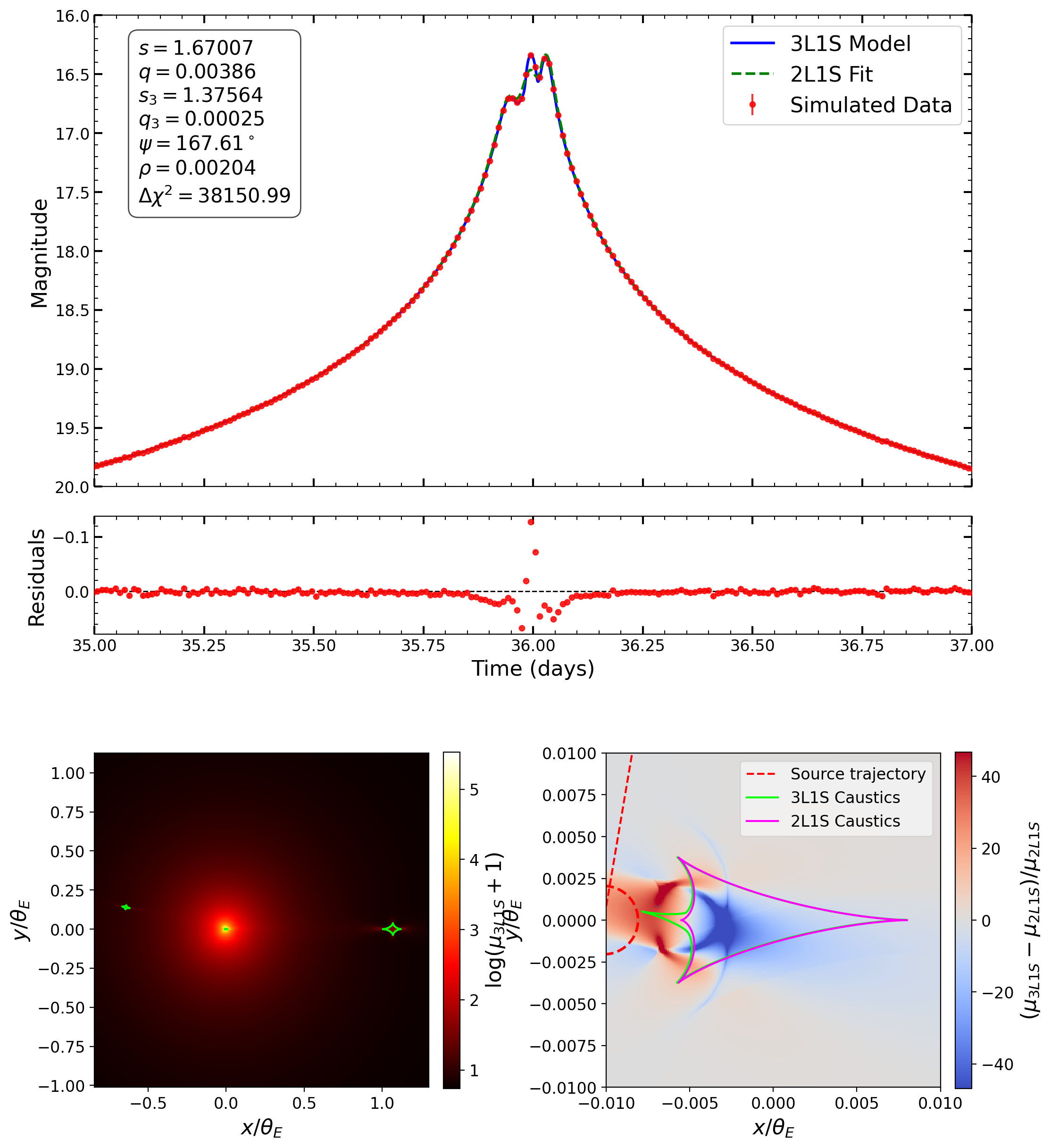}
        \caption{$s$ Wide / $s_3$ Wide}
    \end{subfigure}

    \caption{Simulated light curves for orbital regimes in which at least one planet lies in the wide-separation regime. These configurations generally yield lower detection efficiencies, as planetary caustics are typically detached from the central magnification region.  For each case, the two smaller panels show the triple-lens magnification map and the map of the relative difference between the triple- and binary-lens models near the central caustic.}
    \label{fig:wide_regimes}
\end{figure*}

\begin{figure*}[p]
    \centering
    \begin{subfigure}{0.49\textwidth}
        \includegraphics[width=\linewidth]{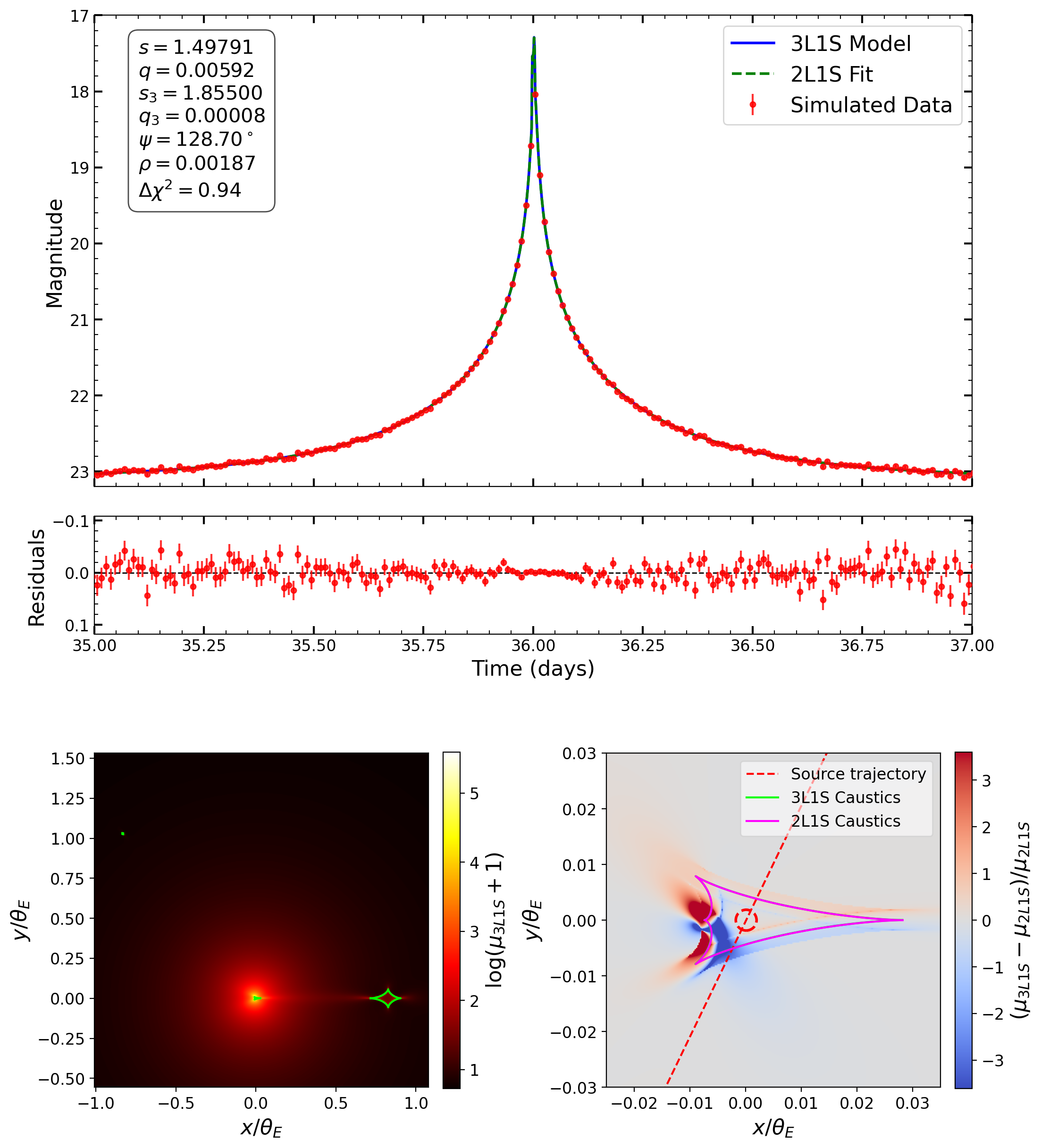}
        \caption{$\Delta\chi^2 < 160$}
    \end{subfigure}
    \hfill
    \begin{subfigure}{0.49\textwidth}
        \includegraphics[width=\linewidth]{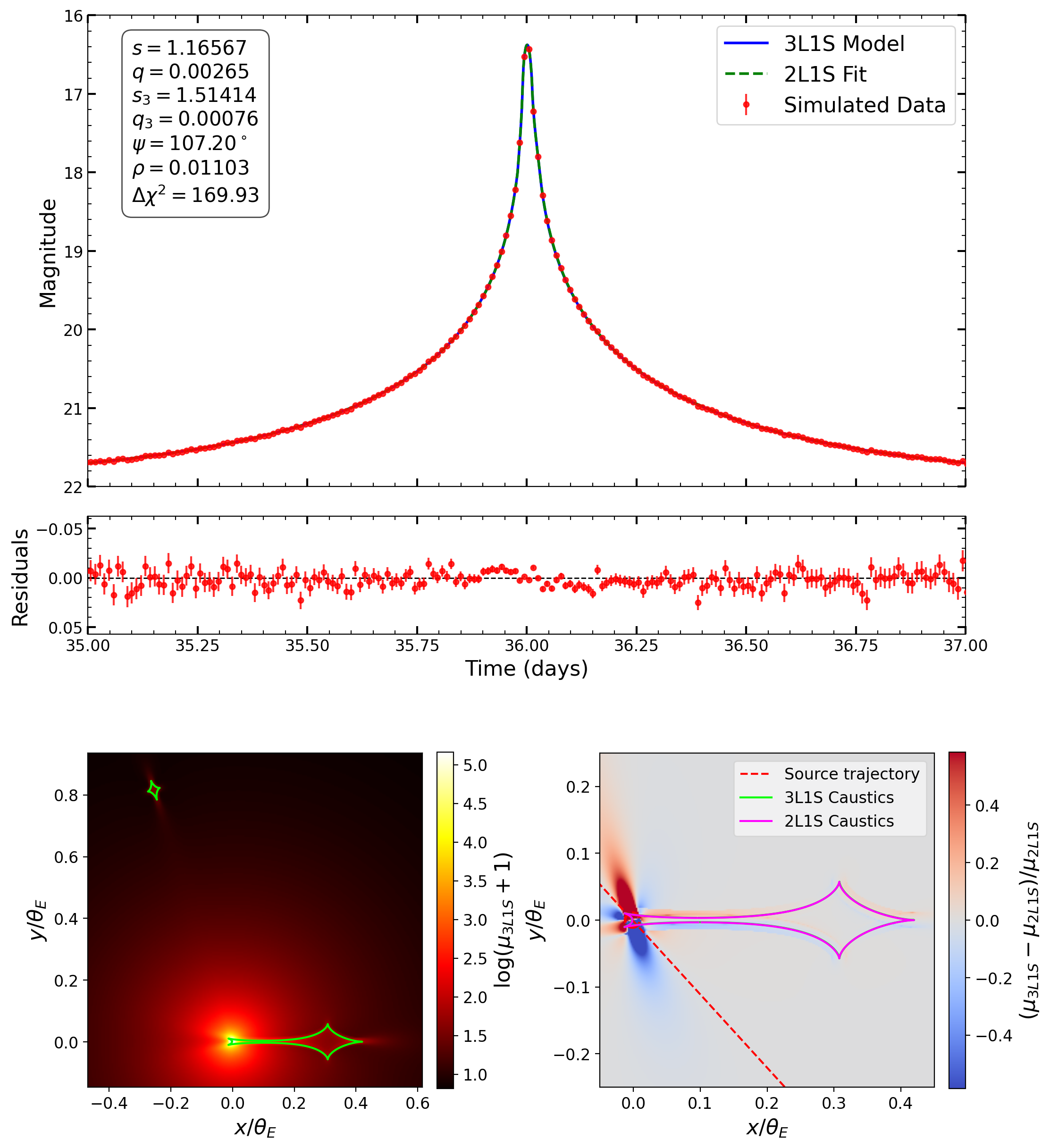}
        \caption{$\Delta\chi^2 \approx 160$}
    \end{subfigure}

    \vspace{1em}

    \begin{subfigure}{0.49\textwidth}
        \includegraphics[width=\linewidth]{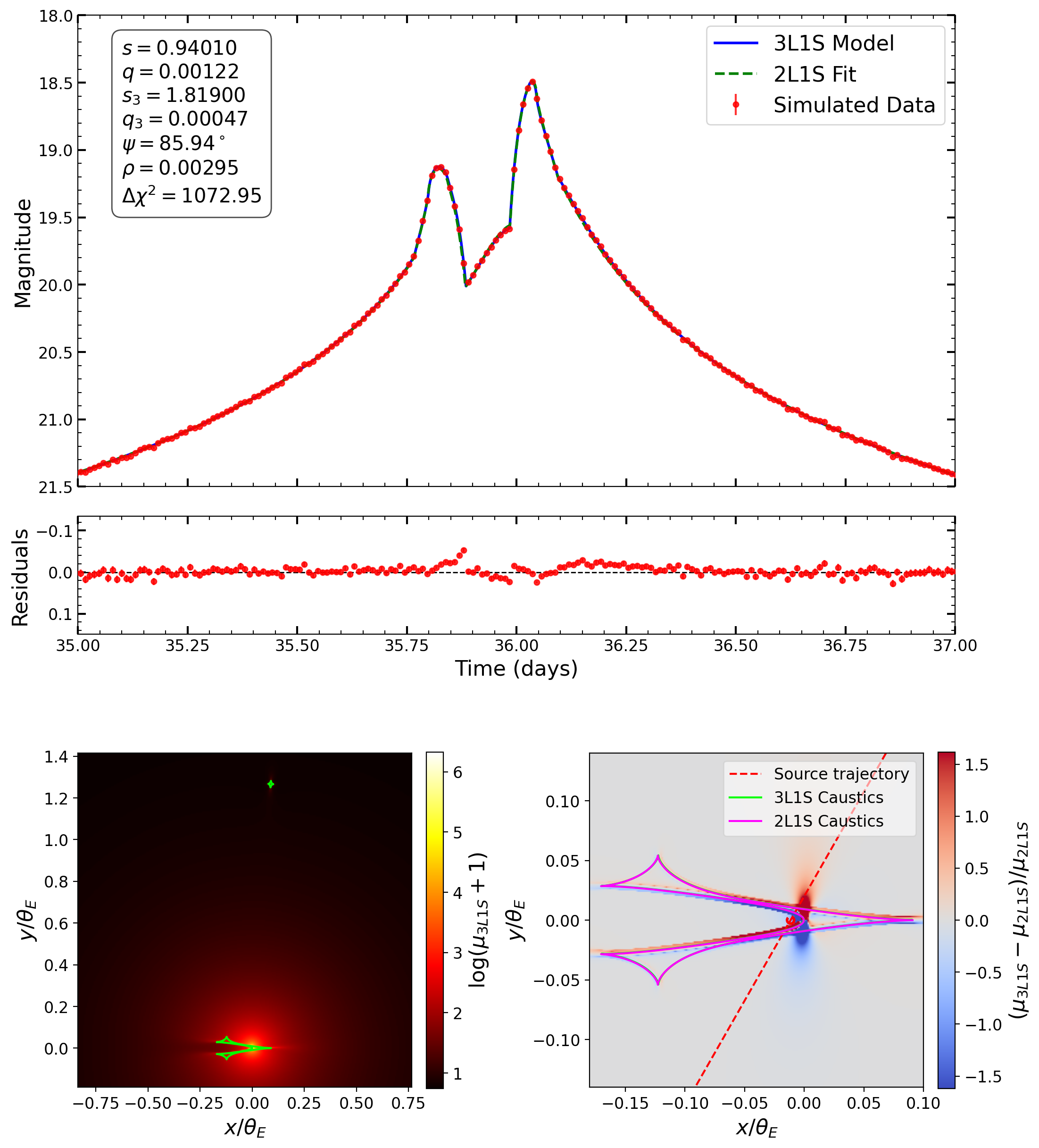}
        \caption{$\Delta\chi^2 > 160$}
    \end{subfigure}

    \caption{Representative simulated microlensing light curves illustrating three cases with increasing detection significance. Panel (a) shows a configuration where the detection threshold of $\Delta\chi^2 = 160$ is not reached, (b) corresponds to a marginal detection near the threshold, and (c) represents a strong detection with $\Delta\chi^2 \approx 1000$.  For each case, the two smaller panels show the triple-lens magnification map and the map of the relative difference between the triple- and binary-lens models near the central caustic.}

    \label{fig:other_cases}
\end{figure*}

Figures~\ref{fig:resonant_regimes}–\ref{fig:wide_regimes} present representative simulated microlensing light curves for all nine orbital configurations defined by the combinations of projected separations $s$ and $s_3$, where $s_3$ corresponds to the less massive planet. Although these examples are not statistically representative, they offer insight into how the geometric configuration of a two-planet system influences the morphology and strength of the resulting anomaly. Figure~\ref{fig:other_cases} illustrates three additional cases: one in which the detection threshold of $\Delta\chi^2 = 160$ is not reached, one that marginally satisfies the threshold, and one displaying a $\Delta\chi^2$ of approximately 1000.
Configurations involving a planet near the Einstein radius (\textit{resonant regime}) typically exhibit prominent, often complex deviations centered around the peak magnification. These arise from large, centrally located caustics and frequently produce blended or multi-feature anomalies that may obscure the individual planetary contributions.
In contrast, systems where both planets lie in non-resonant regimes (e.g., \textit{Close/Close} or \textit{Wide/Wide}) generate simpler, more isolated perturbations. These may appear as short-lived bumps or dips—often weaker and farther from the peak—due to smaller and spatially separated caustics.
Mixed configurations (e.g., \textit{Close/Resonant}, \textit{Wide/Resonant}) often produce two temporally distinct features, one stronger anomaly near the peak from the resonant planet, and a secondary, typically subtler signal from the non-resonant companion. The separation in time and morphology between the two features enhances the likelihood of identifying both planets individually.
Overall, the figure highlights how the presence of a resonant-separation planet tends to dominate the light curve, while the lower-mass companion at $s_3$ contributes more localized or asymmetric perturbations. These patterns help explain the varying detectability of different orbital regimes discussed in the previous section, and provide intuitive guidance for interpreting real multiple-planet microlensing events.

\subsection{Physical Implications of Detection Efficiency Trends}\label{sec:trends-analysis}

We begin by evaluating the effect of the source-lens impact parameter $u_0$ on the detectability of triple-lens events. As shown in Table~\ref{tab:u0_detection_stats}, the detection rate remains broadly stable across the tested values of $u_0$, with no significant trend emerging. The percentage of events exceeding the standard detection thresholds ($\Delta\chi^2 > 160$ and $\Delta\chi^2 > 1000$) varies only mildly, with a slight increase observed for $u_0 < 0.01$.

\begin{table}[ht]
\caption[]{Detection statistics for different $u_0$ values}
\centering
\begin{tabular}{lcc}
\hline \hline
$u_0$ &
Events with &
Events with \\
&
$\Delta\chi^2 > 160$ (\%) &
$\Delta\chi^2 > 1000$ (\%) \\
\hline
0.0001 & 66.23 & 50.35 \\
0.0010 & 69.17 & 53.06 \\
0.0100 & 63.44 & 46.59 \\
\hline
\end{tabular}
\label{tab:u0_detection_stats}
\end{table}

To better understand how physical and geometric parameters affect detectability, we examine the histograms in Figure~\ref{fig:success_rate_histograms}. 
The detectability of a second planet in a microlensing event is found to depend on the planet’s mass ratio, with only negligible sensitivity to the planets’ relative angular orientation. In particular, the histograms in Figure \ref{fig:success_rate_histograms} shows that increasing the outer planet’s mass ratio $q_3$  boosts the triple-lens detection efficiency, whereas varying the mutual angular separation $\psi$ between planets has little systematic effect. As expected, a larger planet-to-star mass ratio produces proportionally larger perturbations in the lens potential, yielding more extended caustics and higher magnification deviations. Consequently, massive companions (e.g. $q \gtrsim 10^{-3}$, on the order of a Jupiter/Sun ratio) generate anomalies that rise well above the photometric noise, leading to a high detection probability. By contrast, a low-mass planet ($q \sim 10^{-4}$ or below) induces only subtle caustic perturbations that often remain buried in noise and thus go undetected. The near-flat response with respect to $\psi$ indicates that rotating the planet configuration hardly changes the total “anomaly cross-section” presented to random source trajectories. Geometrically, the overall area of the combined caustics (and thus the chance that a source will intersect a caustic) is almost invariant under rotating the lens system in the sky. Any modest changes in caustic shape due to different angles do not translate into a strong net difference in detection efficiency.

Another important factor is the Einstein crossing time $t_E$. As seen in the $t_E$ panel, detection efficiency increases with event duration, especially at the short-timescale end. Events with low $t_E$ are more likely to be missed due to poor sampling or blending with noise, whereas longer events allow for better coverage of potential anomalies and a higher probability that at least one perturbation is well sampled.

A similar trend is observed with the parameter \( u_0/\rho \). The detection efficiency increases with this ratio, because larger values correspond to smaller source radii \( \rho \) (for fixed \( u_0 \)), which reduce finite-source smoothing. Smaller sources produce sharper caustic-crossing signatures, making the anomalies more prominent and easier to detect. Conversely, extended sources tend to smooth out the caustic features, lowering the chance of crossing a region of high magnification deviation. This behavior is consistent with what is observed in binary-lens systems, where the finite-source effect can suppress the characteristic signatures of binarity, reducing the detectability of such systems when \( \rho \) is large \citep{2017AJ....154..203B}.

As for the source trajectory angle $\alpha$, detection efficiency appears roughly constant across its full range. This flat trend reflects the near-isotropy of the magnification pattern in high-magnification events: since all considered source paths pass close to the lens center, the direction of approach has little effect on whether anomalies occur, though it may influence their shape. However, because our sample is constructed from previously detectable binary-lens events, the underlying caustic geometry tends to favor trajectories orthogonal with the binary axis, comparable to \cite{2025arXiv250721309B}. This intrinsic lens symmetry can naturally lead to a bimodal distribution in $\alpha$ among the events that show detectable features.

Figure \ref{fig:success_rate} further shows  that detection efficiency increases  with the mass ratios of \emph{both} planets. When either planet is more massive, it contributes a sizeable distortion to the lensing pattern, increasing the likelihood that at least one detectable anomaly will occur during the event. In the most favorable cases—when \emph{both} companions are of high mass (for example, two gas giants with $q, q_3 \gtrsim 10^{-3}$)—our simulations show that nearly all such events are identifiable as triple lenses. By contrast, if both planets are in the sub-$10^{-4}$ mass-ratio regime, the chance of recognizing the second planet is much lower, since each produces only weak perturbations that can be masked by finite-source effects or observational noise.

\begin{figure}
    \centering
    \includegraphics[width=\columnwidth]{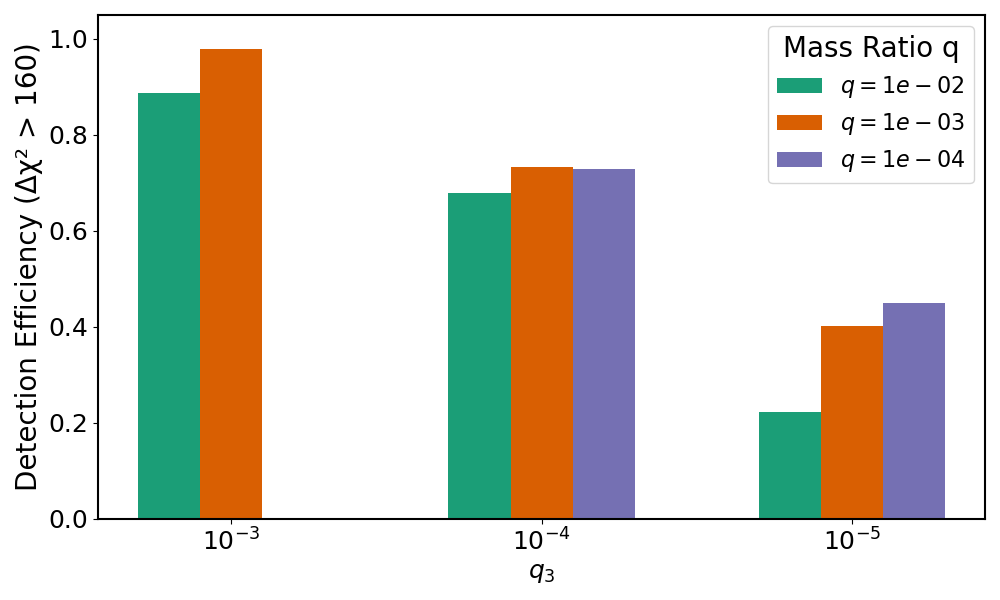}
       \caption{Detection efficiency of a triple-lens system as a function of the mass ratio of the second planet ($q_3$). The efficiency is plotted for three different values of the first planet's mass ratio ($q=10^{-2}$, $q=10^{-3}$, and $q=10^{-4}$). The plot shows that when the second planet is massive ($q_3=10^{-3}$), the detection efficiency is high regardless of the first planet's mass. However, as $q_3$ decreases, the efficiency drops, and this effect is more pronounced when the first planet is more massive.}

    \label{fig:histogram_q_fix}
\end{figure}

The Figure \ref{fig:histogram_q_fix} shows that when the second planet is relatively massive ($q_3 = 10^{-3}$), detection efficiency is nearly 100\% regardless of the primary. As $q_3$ decreases, efficiency drops across all cases, but more steeply when the primary is more massive. Specifically, for $q = 10^{-2}$, efficiency at $q_3 = 10^{-5}$ is significantly lower than for smaller primary planets. This is because a strong primary caustic can overshadow weak perturbations from the secondary, reducing their detectability despite being present.

As evident from Figure~\ref{fig:success_rate_histograms} and also from the Figure~\ref{fig:success_rate}, and consistent with the trends discussed in Section~\ref{sec:orbi_regimes}, a key factor influencing detectability is the separation of the planets relative to the Einstein ring. Our results confirm that the probability of detecting triple systems is highest when at least one planet lies in the resonant regime ($s \sim 1$). This configuration enhances detectability by producing more distinguishable perturbations in the light curve.

In summary, the detection of triple-lens systems is primarily driven by large planet mass ratios and favorable caustic configurations. The highest efficiency is achieved when at least one planet lies near the Einstein ring and has a mass ratio $q \gtrsim 10^{-3}$. In contrast, low-mass companions, extended source sizes (large $\rho$), or short-duration events (low $t_E$) tend to produce weak or smoothed-out signals that often fall below the detection threshold. Overall, planetary mass, source size, event duration, and resonant geometry play a dominant role in shaping the success rate, while other factors such as the trajectory angle $\alpha$ or mutual angular orientation $\psi$ have only a minor impact on detectability.
\section{Conclusions}\label{sec:Conclusions}

We have assessed the capability of the \textit{Nancy Grace Roman Space Telescope} to detect and characterize triple-lens microlensing events—planetary systems in which two exoplanets orbit a common host star. By analyzing a large set of simulated high-magnification microlensing light curves, incorporating realistic photometric noise and Roman’s anticipated survey parameters, we quantified the conditions under which such triple-planet systems can be reliably distinguished from simpler binary-lens (single-planet) models.

Our results indicate that 66.3\% of simulated high-magnification triple-lens events would be identifiable under a demanding detection threshold ($\Delta\chi^2 > 160$), demonstrating robust sensitivity to multiple-planet signatures. In favorable configurations where both planets have relatively large mass ratios (on the order of $10^{-3}$), the detection success rises above 90\%, whereas systems with both planets in the low-mass regime ($\sim 10^{-4}$) yield significantly lower detection rates. These trends highlight Roman's potential to uncover a substantial population of multi-planet systems through microlensing, improving significantly upon the presently scarce detections of multiple planets by this method.

The detectability of triple-lens systems is governed primarily by the planetary mass ratios and the event geometry—in particular, the source star’s trajectory relative to the lens’s caustic structure. If the source path intersects or closely approaches the combined central caustic generated by a two-planet configuration, the resulting light-curve deviations cannot be replicated by any binary-lens model, providing a clear indication of an additional planet.

We find that the specific planetary configuration plays an important role: resonant orbital separations, particularly when the more massive planet lie near the Einstein ring, produce extended and overlapping central caustics that dominate the high-magnification region, yielding the highest detection efficiencies. In contrast, configurations in which both planets are well inside or outside the Einstein ring (e.g., close–close or wide–wide regimes) produce smaller, offset caustics that are less likely to be intersected by the source, leading to substantially lower detection probabilities.

Thus, high planetary mass ratios and resonant star–planet separations define the most favorable conditions for Roman to reveal additional planets via microlensing. Widely separated, low-mass systems with unfavorable geometry, on the other hand, are more likely to remain undetected.

Relying on the fraction of high-magnification events and our simulated detection efficiencies, we estimate that the probability of detecting a triple-lens microlensing event with \textit{Roman} is approximately $0.1\%$, corresponding to around 60 events over the course of the survey. While modest, this estimate represents a significant improvement over ground-based capabilities and a promising step toward characterizing complex planetary systems.

In summary, Roman’s microlensing survey will expand the discovery space for multiplanetary systems. Roman not only will increase the detection rate of individual wide-separation exoplanets, but also will play a pivotal role in identifying systems hosting multiple planets, opening a new observational window onto planetary architectures that are inaccessible to transit and radial-velocity techniques.

Future work: Building on these results, we plan to extend our simulations to explore a broader parameter space—including varying source star radii, orbital orientations of the planets, and second-order effects such as microlensing parallax and lens orbital motion—in order to refine detectability forecasts for triple-lens systems under more diverse conditions.

\bibliographystyle{aa} 
\bibliography{bibliography}

\begin{acknowledgements}
      This work was carried out during a visiting period at IPAC, California Institute of Technology, as part of the activities of the Roman  Project Infrastructure Team (PIT). VS warmly thanks the entire IPAC team for their hospitality and stimulating environment.\\
      
      VS and GC acknowledge support by the University of Napoli Federico II (project: FRA-CosmoHab, CUP E65F22000050001) and thank the university for facilitating the visiting period. \\
      
      VB acknowledges financial support from PRIN2022 CUP D53D23002590006.\\

      Funding for the Roman Galactic Exoplanet Survey Project Infrastructure Team is provided by the Nancy Grace Roman Space Telescope Project through the National Aeronautics and Space Administration grant 80NSSC24M0022, by The Ohio State University through the Thomas Jefferson Chair for Space Exploration endowment, and by the Vanderbilt Initiative in Data-intensive Astrophysics (VIDA).\\

      SIS’s research was supported by an appointment to the NASA Postdoctoral Program at the NASA Goddard Space Flight Center, administered by Oak Ridge Associated Universities under contract with NASA.\\
      
      This research was supported in part by computational resources provided by the NASA High‑End Computing (HEC) Program through the NASA Advanced Supercomputing (NAS) Division at Ames Research Center, including use of the Fornax Science Console platform developed by NASA’s Astrophysics archives (IRSA, MAST, HEASARC) We gratefully acknowledge NASA for access to these resources, which made the completion of this work possible. \\

      We acknowledge the use of the ADHOC (Astrophysical Data HPC Operating Center) resources, within the project "Strengthening the Italian Leadership in ELT and SKA (STILES)", proposal nr. IR0000034, admitted and eligible for funding from the funds referred to in the D.D. prot. no. 245 of August 10, 2022 and D.D. 326 of August 30, 2022, funded under the program "Next Generation EU" of the European Union, “Piano Nazionale di Ripresa e Resilienza” (PNRR) of the Italian Ministry of University and Research (MUR), “Fund for the creation of an integrated system of research and innovation infrastructures”, Action 3.1.1 "Creation of new IR or strengthening of existing IR involved in the Horizon Europe Scientific Excellence objectives and the establishment of networks".
\end{acknowledgements}
\label{LastPage}
\end{document}